%% file: ejn_paper.tex
\documentclass[alpha-refs]{wiley-article}
\usepackage{graphicx}
\usepackage{mwe}
\usepackage[space]{grffile}
\usepackage{latexsym}
\usepackage{textcomp}
\usepackage{longtable}
\usepackage{tabulary}
\usepackage{booktabs,array,multirow}
\usepackage{amsfonts,amsmath,amssymb}
\usepackage{natbib}
\usepackage{url}
\usepackage{hyperref}
\hypersetup{colorlinks=false,pdfborder={0 0 0}}
\usepackage{etoolbox}

\newif\iflatexml\latexmlfalse

\AtBeginDocument{\DeclareGraphicsExtensions{.pdf,.PDF,.eps,.EPS,.png,.PNG,.tif,.TIF,.jpg,.JPG,.jpeg,.JPEG}}

\usepackage[utf8]{inputenc}
\usepackage[english]{babel}

\usepackage{float}

\usepackage{makecell}
\usepackage{anyfontsize}
\usepackage{adjustbox}
\usepackage{xcolor}
\usepackage[normalem]{ulem}
\usepackage{ifthen}
\usepackage{placeins}
\usepackage[most]{tcolorbox}
\usepackage{etoolbox}

\newif\ifrevision
\revisionfalse 

\newcommand{\rev}[1]{%
  \ifrevision
    {\color{black}#1}%
  \else
    #1%
  \fi
}

\iflatexml

\else

\paperfield{ORIGINAL ARTICLE FOR CONSIDERATION IN EJN: FREEZING OF GAIT - A TRIBUTE TO NIR GILADI}
\abbrevs{FOG, freezing of gait; PD, Parkinson's disease; HAR, human activity recognition; ML, machine learning.}
\corraddress{Department of Biomedical Informatics, Emory University, Atlanta, GA, USA}
\corremail{hyeokhyen.kwon@emory.edu}
\presentadd{Department of Biomedical Informatics, Emory University, Atlanta, Georgia, USA}
\fi

\papertype{Special Issue Article}
\title{Evidence for Phenotype-Driven Disparities in Freezing of Gait Detection and Approaches to Bias Mitigation}

\author[1]{Timothy Odonga}
\author[1]{Jeanne M. Powell}
\author[1]{Claire Bae}
\author[1]{Yumeng Tan}
\author[2]{Christine D. Esper}
\author[2]{Stewart A. Factor}
\author[1,2,3]{J. Lucas Mckay\textsuperscript{\textdagger}}
\author[1,3]{Hyeokhyen Kwon\textsuperscript{\textdagger}}

\affil[1]{Department of Biomedical Informatics, School of Medicine, Emory University, Atlanta, Georgia, USA}
\affil[2]{Jean and Paul Amos Disease and Movement Disorders Program, Department of Neurology, School of Medicine, Emory University, Atlanta, Georgia, USA}
\affil[3]{Wallace H. Coulter Department of Biomedical Engineering, Georgia Institute of Technology, Atlanta, Georgia, USA}

\presentadd{\textdagger These authors contributed equally to this work as senior authors.}

\runningauthor{Odonga et al.}

\newboolean{highlighted}
\setboolean{highlighted}{false} 

\ifthenelse{\boolean{highlighted}}{
\usepackage{lineno}
\newcommand{\add}[1]{\textcolor{blue}{#1}}
\newcommand{\del}[1]{\textcolor{red}{\sout{#1}}}

\newcommand{\addBox}[1]{%
    \begin{tcolorbox}[
        colback=white,
        colframe=blue,
        boxrule=0.8pt,
        sharp corners,
        left=2pt,
        right=2pt,
        top=2pt,
        bottom=2pt
    ]
    #1
    \end{tcolorbox}
}

\newcommand{\addLong}{\color{blue}\arrayrulecolor{blue}}

}{

\newcommand{\add}[1]{\textcolor{black}{#1}}
\newcommand{\del}[1]{}
\newcommand{\addBox}[1]{#1}
\newcommand{\addLong}{}

}

\begin{document}

\maketitle
\selectlanguage{english}
\begin{abstract}

\input{sections/abstract}
\textbf{Keywords} --- Parkinson's Disease, Wearable Sensors, Human Activity Recognition
\end{abstract}

\ifthenelse{\boolean{highlighted}}{
\linenumbers
}{
}

\input{sections/intro}

\input{sections/methods}

\input{sections/results}

\input{sections/discussion}

\input{sections/conclusion}

\selectlanguage{english}

\clearpage
\add{\bibliography{ref}}

\clearpage
\appendix
\input{sections/supplementary}

\end{document}

%% file: sections/abstract.tex
\par Freezing of gait (FOG) is a debilitating feature of Parkinson's disease (PD) and a common cause of injurious falls. 
Recent advances in wearable-based human activity recognition (HAR) have enabled FOG detection, but bias and fairness in these models remain understudied. 
Bias is defined as a systematic error leading to unequal outcomes, while fairness refers to consistent performance across subject groups. 
Biased models could systematically underserve patients with specific FOG phenotypes or demographics, widening care disparities.
We systematically evaluated the bias and fairness of state-of-the-art HAR models for FOG detection across FOG phenotypes and patient demographics using multi-site datasets. 
\add{We used 4 public wearable datasets (Total of 145 subjects and  59.8 hours of recording) with multiple on-body sensor locations (ankle, lower back, thigh, and shank) and temporal annotations on FOG episodes with various behavior test protocols, including Timed Up \& Go, turning, sitting and rising from chairs, etc.}
Four mitigation approaches were assessed: conventional methods (threshold optimization and adversarial debiasing) and two transfer learning-based approaches (multi-site transfer and fine-tuning pretrained models). 
Fairness was primarily quantified using the demographic parity ratio (DPR) and the equalized odds ratio (EOR).
HAR models exhibited substantial bias (DPR \& EOR < 0.8) across age, sex, disease duration, and critically, FOG phenotype. 
This phenotype-specific bias is particularly concerning as tremulous and akinetic FOG may require different clinical management.
Conventional methods failed: threshold optimization (DPR=$-0.126$, EOR=$+0.063$) and adversarial debiasing (DPR=$-0.008$, EOR=$-0.001$). 
In contrast, multi-site transfer learning \del{significantly} improved fairness (DPR=+0.037, p<0.01; EOR=+0.045, p<0.01) and performance (F1-score=
+0.020, p<0.05)\add{, while not meeting the fairness threshold of DPR>0.8 and EOR>0.8}.
Transfer learning across diverse datasets \add{may partially mitigate disparities}\del{is essential} for developing HAR models that reliably detect FOG across all patient phenotypes, ensuring wearable-based monitoring benefits all individuals with PD.

%% file: sections/intro.tex
\section{Introduction}

\par Freezing of gait (FOG) is a debilitating feature of Parkinson's disease (PD) that is difficult for clinicians to capture and evaluate during time-limited clinical assessments~\citep{nutt2011freezing}.
This is primarily due to its brief, sudden, and episodic nature, which makes it challenging to assess on demand~\citep{nutt2011freezing,gilat2026updated}. 
Furthermore, clinical assessments may not fully capture FOG as subjects’ gait typically improves when they concentrate on walking during evaluation~\citep{hallett2008intrinsic}, and when they are assessed in typically well-lit hospital environments~\citep{barthel2016practicalities}. 
In addition, subjects are often evaluated when in an on-medication state that has less occurrence of FOG in comparison to the off-medication state~\citep{schaafsma2003characterization, mckay2019persist}. 
Consequently, it is important to assess FOG outside of routine clinical visits in order to more accurately determine its impact on activities of daily living~\citep{mancini2019clinical}.


\par Machine learning approaches using wearable or kinematic data show promise for objectively detecting FOG and enabling automated assessment~\cite{kwon2023, filtjens2022automated, mancinimckay2025, salomon2024contest, yang2024automatic}.
Studies have shown that wearable sensors can provide objective FOG assessment with sensitivity and specificity values exceeding 90\% across different sensor configurations, provoking tasks, detection algorithms, and validation settings~\citep{reches2020using, pardoel2019wearable}.
Recent studies have also shown that kinematic marker data from optical motion capture systems enable automated FOG assessment, with deep learning approaches achieving high accuracy and identifying features consistent with expert assessments~\citep{filtjens2022automated, kwon2023}.


\par A complicating feature of FOG is that it manifests in distinct phenotypes, such as kinetic (mostly tremulous) and akinetic forms, which may reflect different underlying pathophysiology and treatment needs~\citep{factor2025possible, gilat2026updated}.
Akinetic FOG involves impairments across motor, cognitive, and limbic basal ganglia circuits and emerges predominantly during gait initiation and dual-tasking, while tremulous FOG reflects relatively preserved cognitive and limbic function with unsuccessful stepping attempts that predominate during turning~\citep{koehler2021freezing, zoetewei2025effects}. 
Phenotype-specific detection challenges may thus contribute to systematic differences in automated FOG assessment.


\par Few of these models have been comprehensively evaluated across phenotypes, which may introduce systematic error. 
Best practices for assessing differential performance have yet to be established. 
Few prior studies have reported phenotype-stratified performance; limited evidence from one study~\citep{yang2024automatic} suggests higher performance for tremulous compared to akinetic FOG (intraclass correlation coefficient [ICC] roughly 0.86 vs. 0.78). 
However, that study focused on optimizing overall detection accuracy rather than systematically evaluating performance across phenotypes. 
Performance was not examined across subject groups, nor were potential contributors to reduce accuracy in akinetic FOG explored. 
Given that tremulous and akinetic manifestations co-occur in over 80\% of patients~\citep{zoetewei2025effects}, systematic evaluation of performance differences across FOG manifestations remains an important gap.


\par At the same time, the machine learning community has established methods to quantify bias in predictive models and mitigate it through fairness methods.
Bias is defined as a systematic error that leads to unequal outcomes~\citep{ferrara2023fairness}, while fairness is defined as consistent performance across subject groups and phenotypes~\citep{Varshney2021}.
Bias can be quantified across subject groups at multiple levels, including metrics based solely on predicted outcomes and metrics that compare predicted outcomes against actual labels~\citep{verma2018fairness}.
Bias mitigation methods (fairness methods) can be applied at different stages of the modeling pipeline: pre-processing methods address imbalanced training data, in-processing methods incorporate fairness constraints into model optimization~\citep{Zhang2018}, and post-processing methods adjust decision thresholds to equalize performance across groups~\citep{Hardt2016}. 
These techniques have been successfully applied to reduce performance disparities in settings such as clinical risk prediction and diagnostic imaging algorithms~\citep{sivarajkumar2023fair, yang2023algorithmic, siddiqui2024fair}. 
However, these established fairness methods have not been systematically applied to FOG detection models, so it remains unclear whether they can effectively mitigate differential performance across FOG manifestations.


\par To the best of our knowledge, no prior work has applied formal fairness in machine learning (Fair ML) frameworks to investigate bias and fairness in wearable-based FOG detection, despite their importance for developing unbiased models that measure FOG accurately across phenotypes or demographic strata.
While a few studies in FOG and PD research have reported performance differences across subject groups, these analyses were limited to post hoc statistical comparisons and did not employ Fair ML techniques to address the identified differences. 
In FOG detection, \citet{kwon2023} computed individual-level F1 scores for each participant and compared performance across age, sex, and study groups using linear regression with Wald tests for statistical significance. 
In broader PD digital biomarker research,~\citet{zhang2020deep} performed stratified analyses by calculating performance metrics with 95\% confidence intervals for each demographic stratum and conducting pairwise bootstrap comparisons, while~\citet{khera2022age} trained separate prediction models for different age-gender groups and compared classification accuracies using 10-fold cross-validation. 
\citet{adnan2025ai} reported performance differences across sex and ethnic subject groups for PD screening from facial videos.
However, neither study quantified bias using Fair ML metrics or applied bias mitigation strategies, and wearable-based FOG detection remains entirely unexamined across phenotypic and demographic subject groups.



\par To address this gap, we implemented five existing state-of-the-art human activity recognition (HAR) models~\cite{breiman2001random, ordonez2016,haresamudram2020, yuan2024self, ruan2025foundationmodelswearablemovement} for FOG detection and systematically evaluated their bias and fairness across FOG phenotypes and demographic strata. 
We used four multi-site datasets to make our approach as representative as possible~\cite{misc_daphnet_freezing_of_gait_245, boari2022, kaggleFOG}. 
Models were implemented either by deploying open-source code from original publications or by replicating published architectures. 
We used three common notions of group fairness to comprehensively evaluate the FOG detection models~\cite{verma2018fairness}. 
Demographic parity assesses whether positive prediction rates are equal across subject groups~\cite{Dwork2012}. 
Equalized odds require equal true positive and false positive rates across subject groups conditional on the true outcome~\cite{Hardt2016}. 
Equal opportunity, a relaxation of equalized odds, requires only equal true positive rates across subject groups~\cite{Hardt2016}. 
These three notions offered complementary perspectives on fairness, assessing both prediction patterns and prediction accuracy across subject groups~\cite{verma2018fairness}.
Lastly, we implemented various bias mitigation approaches, such as threshold optimization~\cite{Hardt2016} and adversarial debiasing~\cite{Zhang2018}, as well as transfer learning-based approaches using multi-site data and large-scale pretrained models to assess the extent to which these approaches would mitigate biases. 

\par Despite high overall accuracy, FOG detection models may mask clinically significant failures in specific subject groups and especially when those subgroups require different treatment strategies~\cite{mitani2021survey, mehrabi2021survey}. 
We hypothesized that models would show systematic differences across FOG phenotypes, with lower detection of akinetic FOG than of tremulous FOG, and that fairness methods could reduce these gaps~\cite{yang2024automatic}. 
Our aims were to: (1) quantify the extent of bias in current FOG detection models across phenotypes and demographics using established Fair ML metrics, and (2) evaluate whether the various bias mitigation approaches implemented can achieve fairness across phenotypes and demographics.

%% file: sections/methods.tex
\section{Methods}

\subsection{Overall Approach}
Here, we describe the definitions and metrics of bias and fairness used in our analysis, the four multi-site datasets, the HAR models implemented, the bias mitigation approaches, and the experimental settings used to evaluate FOG detection performance and model biases in this work.
\autoref{fig:fairml_pipeline} shows the overall framework of our evaluation.
\input{figures/fair_ml_pipeline}

\subsection{Definitions of Bias and Fairness}
In this section, we define protected attributes, bias, and fairness. 
We then describe the notions of group fairness evaluated in this study and the metrics used to quantify them.

\par Protected attributes are attributes of an individual that may not be used directly or indirectly as a basis for automated decision-making in order to prevent systematic advantage accorded to specific groups in automated decision-making processes~\cite{Dwork2012, Varshney2021}.
While no fixed set of protected attributes exists, this work considers clinical attributes, such as FOG phenotype and disease duration, as well as demographic characteristics, such as age and sex. 
Model performance should be consistent across groups defined by these attributes, ensuring that model predictions are independent of these attributes~\cite{Dwork2012, Varshney2021}.

Bias in automated decision support systems is defined as systematic errors leading to unequal outcomes~\cite{ferrara2023fairness}. 
Fairness can be defined at the individual level (similar treatment for similar individuals) or at the group level (equal behavior across demographic groups)~\cite{Dwork2012, Varshney2021}. 
This work adopts a group fairness definition, whereby fairness is satisfied when average model performance is consistent across groups defined by protected attributes (e.g., FOG phenotype, age, sex)~\cite{Varshney2021}.

\par Demographic parity requires equal proportions of positive predictions between protected groups (e.g., equal rate of detected FOG episodes from the model)~\cite{Dwork2012}.
In this work, demographic parity was computed using the \textbf{Demographic Parity Ratio (DPR)} defined below~\cite{weerts2023fairlearn}.
\begin{equation}
    DPR = \frac{P(F_{pred} = 1 | G = g_i)}{P(F_{pred} = 1 | G = g_j)}
    \label{eq:dpr}
\end{equation}
where $F_{pred}$ is the model's detection of a FOG episode, $i \neq j$, $g_i$ and $g_j$ are binary groups derived from a protected attribute G and $P(F_{pred} = 1 | G = g_i) \leq P(F_{pred} = 1 | G = g_j)$.

Equality of opportunity (true positive parity) is satisfied when protected groups benefit equally from correct model predictions (e.g., a model correctly detects an FOG episode)~\cite{Hardt2016}. 
Conversely, false positive parity requires equal rates of incorrect positive predictions across groups (e.g., a model detects a false FOG episode)~\cite{Hardt2016}.
True positive parity ratio (TPPR) and false positive parity ratio (FPRR) are defined as shown below~\cite{Hardt2016}.
\begin{equation*}
    TPPR = \frac{P(F_{pred} = 1 | F_{true} = 1, G = g_i)}{P(F_{pred} = 1 | F_{true} = 1, G = g_j)}
\end{equation*}
\begin{equation*}
    FPRR = \frac{P(F_{pred} = 1 | F_{true} = 0, G = g_i)}{P(F_{pred} = 1 | F_{true} = 0, G = g_j)}
\end{equation*}
where $F_{pred}$ is the model's output, $F_{true}$ is the ground truth annotation for a FOG episode, $i \neq j$, $g_i$ and $g_j$ are binary groups derived from a protected attribute G and $P(F_{pred} = 1 | G = g_i) \leq P(F_{pred} = 1 | G = g_j)$.
Equality of opportunity was computed using the \textbf{Equality of Opportunity Difference (EOD)}~\cite{weerts2023fairlearn}:
\begin{equation}
    EOD = P(F_{pred} = 1 | F_{true} = 1, G = g_i) - P(F_{pred} = 1 | F_{true} = 1, G = g_j)
     \label{eq:eod}
\end{equation}

Equalized odds are satisfied when the true positive parity and the false positive parity are equal across groups~\cite{Hardt2016}.
Equalized odds were computed using the \textbf{Equalized Odds Ratio (EOR)}, which captures the worst-case disparity between groups, as shown below~\cite{Hardt2016, weerts2023fairlearn}.
\begin{equation}
    EOR = min(TPPR, FPRR)  
    \label{eq:eor}
\end{equation}

\subsection{Wearable Datasets for FOG}
\label{sec:dataset}

\input{tables/dataset_details_reviews}

To effectively evaluate FOG detection performance across phenotypic and demographic groups, four publicly available multi-site FOG datasets were selected based on variation in sensor locations, FOG-provoking protocols, and the location of clinical assessment (i.e., clinic versus home).
Across datasets, there were a total of N=145 individuals with PD, of whom $31\%$ were women, the average age was $67\pm9$ years, and the average disease duration was  $10\pm6$ years.
All of the datasets used common inertial measurement unit (IMU) instrumentation with varied sensor locations across the lower back, ankle, shank, and thigh. 
All of them exhibited a significant label imbalance with No FOG predominating the recorded samples ($86\%$ on average) and tremulous FOG much more prevalent than akinetic FOG ($82\%$ on average).
Importantly, these datasets contain sample-level FOG annotations enabling extraction and phenotypic classification of individual FOG episodes~\cite{moore2013autonomous}.
The specifics of the FOG-provoking protocols varied substantially across datasets, including Timed Up \& Go~\cite{podsiadlo1991timed}, Hotspot Door tasks~\cite{kaggleFOG}, motor cognitive tasks~\cite{reches2020using}, daily walking tasks~\cite{daphnet_paper}, and turning tasks~\cite{boari2022}.
These details are summarized in~\autoref{tab:dataset_protocol_details}.
Characteristics of each included dataset are summarized in~\autoref{tab:dataset_details_reviews}.

\add{
All datasets provided binary FOG annotations at the individual-sample level, but information regarding FOG phenotype was limited. The DEFOG and tDCS FOG datasets additionally included labels describing the apparent provoking factor (“StartHesitation,” “Turn,” or “Walking”) and a dichotomous “Kinetic” variable. To generate harmonized phenotype labels across datasets, we first grouped contiguous samples labeled as FOG into discrete episodes. We then applied the spectral method described by Moore et al~\cite{moore2013autonomous}, to classify each episode as putatively tremulous or akinetic. Specifically, episodes in which accelerometer power in the freezing band (3–8 Hz) exceeded power in the locomotor band (0–3 Hz) were classified as tremulous and all remaining episodes were classified as akinetic. Where expert-provided “Kinetic” labels were available, we evaluated agreement between the derived and expert labels, treating our tremulous label as corresponding to an expert “Kinetic” value of 1 and our akinetic label as corresponding to an expert “Kinetic” value of 0. Overall classification accuracy was 79.6\%, which we considered acceptable for the planned phenotype-stratified analyses (See \autoref{fig:phenotype_annotation_confusion_matrix}). However, this comparison represents an imperfect validation because the expert-provided “Kinetic” category may encompass tremulous FOG as well as other kinetic FOG manifestations.
}

\subsection{Pipeline and HAR Models}



\par To effectively capture FOG episodes and examine bias in the state-of-the-art HAR models, we first processed our data using established activity recognition approaches, including the sliding window approach~\cite{bulling2014tutorial}.
Reported optimal window sizes that maximized FOG detection performance were used to segment recordings into non-overlapping windows for FOG detection, namely, 3-second windows for tDCSFOG, DeFOG, and DeSouza~\cite{reches2020using}, and a 4.5-second window for Daphnet~\cite{daphnet_paper}.
Additionally, Min-Max scaling was applied to each sensor channel across all subject readings to accelerate model convergence during training~\cite{Rosati2018}.

We selected five HAR model types to represent the current state of the art while maintaining an element of parsimony~\cite{breiman2001random, ordonez2016,haresamudram2020, yuan2024self, ruan2025foundationmodelswearablemovement}.
The first three HAR models (referred to as \textbf{Models A1-A3}) represent well-established HAR methodologies commonly available in the literature and were trained from scratch: Random Forest (RF; \textbf{Model A1}), a deep convolutional and long short-term memory model (DeepConvLSTM; \textbf{Model A2})~\cite{ordonez2016}, and Masked Transformer (\textbf{Model A3})~\cite{haresamudram2020}.
These models represent a class of architectures commonly used for tasks such as detecting daily activities (e.g., walking, sitting, standing), detecting sports and exercise movements, and gesture recognition~\cite{zappi2008activity, roggen2010collecting}.

Model A1 and A2 were selected based on their known ability to classify complex activities in daily living and locomotion~\cite{kwon2019handling}.
Model A3 was selected because of the ability of transformer-based models to model long-range dependencies and sequence data~\cite{vaswani2017attention}.
For Model A1, we extracted features from each analysis window using the empirical cumulative distribution function (ECDF) to obtain nonparametric, length-invariant representations that preserve the empirical distributions of each sensor channel~\cite{hammerla2013preserving}.
Following Ordóñez and Roggen~\cite{ordonez2016}, the raw sensor time series were provided to the deep learning-based models (Models A2-A3).
To enable real-time prediction of the onset or offset of FOG episodes during subject movement, each analysis window was labeled as having an FOG episode or normal movement based on the annotation at the last timestep of the window~\cite{yao2018efficient}.

We also fine-tuned two large-scale pretrained models (also referred to as foundation models) (\textbf{Models B1-B2}): a HAR-based pretrained model~\cite{yuan2024self} (\textbf{Model B1}) and a foundation model based on actigraphy data~\cite{ruan2025foundationmodelswearablemovement} (\textbf{Model B2}).
In general, these models offer improved recognition in use cases like activity classification, sleep abnormality detection, and mental health biomarker detection~\cite{yuan2024self, ruan2025foundationmodelswearablemovement}. 
They differ from models A1-A3 in that they are pretrained on petabyte-scale data that may provide useful "inherited" features learned during pretraining, potentially improving performance on downstream classification tasks even with limited labeled training data. 
Model B1 is a residual network-based (ResNet) convolutional model pretrained on general human activities (e.g., walking, sitting, cycling) from UK Biobank~\cite{yuan2024self}, while Model B2 is a transformer-based model pretrained on actigraphy data~\cite{ruan2025foundationmodelswearablemovement}.
While ResNet's convolutional layers excel at detecting local patterns but struggle with long-range dependencies, transformers can capture relationships across distant time points through their attention mechanisms~\cite{He2016, vaswani2017attention}.
Similar to Model A2-A3, the raw sensor time series were provided to Models B1-B2~\cite{ordonez2016}.


\subsection{Bias Mitigation}
\par We implemented four bias mitigation approaches spanning the full machine learning pipeline and compared established fairness methods with novel transfer learning approaches to identify which strategies most effectively reduced bias in FOG detection models. 
Two established methods (threshold optimization and adversarial debiasing) have demonstrated effectiveness in fairness applications across other domains such as finance~\cite{Hardt2016, Zhang2018} and mental health~\cite{jiang2024evaluating,mu2026detecting}. 
We additionally evaluated two transfer learning approaches (leveraging multi-site data and pretrained models) that we hypothesized would improve fairness by exposing models to more diverse training examples.
Due to architectural constraints and practical considerations, we applied each mitigation approach only to models for which prior literature suggested optimal performance and technical feasibility.
For example, adversarial debiasing requires gradient-based optimization and therefore could not be applied to Model A1 (Random Forest)~\cite{Zhang2018}.
Similarly, threshold optimization was not applied to the deep learning models (Models A2-A3), as it has been shown to be suboptimal for correcting biases encoded in deep learning models' learned representations~\cite{chen2023algorithm, kpatcha2025balancing}.

\par Threshold optimization, known for its effectiveness with shallow models, was applied to Model A1 to reduce bias quantified by DPR, EOR, and EOD between the protected groups~\cite{Hardt2016}. 
The concept of threshold optimization is a post-processing bias mitigation method that adjusts decision thresholds derived from a model's score distribution, separately for different protected groups, to reduce differences in error or selection rates without retraining the model~\cite{Hardt2016}.
Similar to using sensitivity or specificity to identify an optimal operating point for a generic classifier, threshold optimization selects group-specific thresholds that satisfy fairness criteria, enabling a controlled trade-off between accuracy and group fairness~\cite{Hardt2016}.
Decision thresholds for each protected group were adjusted to satisfy demographic parity, true positive parity, and equalized odds in our implementation~\cite{Hardt2016}.
To mimic real-world deployment scenarios, threshold optimization was applied only to the train, and validation splits to calibrate the decision thresholds, with evaluation conducted on the test data, following a nested cross-validation paradigm~\cite{weerts2023fairlearn, Hardt2016}.

\par We applied adversarial debiasing to train attribute-agnostic feature representations in the implemented FOG detection models~\cite{Zhang2018}.
The concept of adversarial debiasing is that two competing neural networks are trained simultaneously: the main classifier learns to detect FOG episodes while a second "adversarial" model examines the main classifier's internal representations and attempts to identify protected attributes (such as age or sex) from them~\cite{Zhang2018}. 
During training, the main classifier receives a penalty whenever the adversarial model successfully identifies these attributes, forcing it to learn patterns (in this case, FOG patterns) that do not encode demographic or clinical characteristics~\cite{Zhang2018}. 
By combining the main model's performance with the penalty from the adversarial model, this approach aims to create a classifier that performs FOG detection without relying on protected attributes.
This was implemented with two simultaneously trained neural networks: our deep-learning predictor model (Model A2 or A3) and a two-layer multi-layer perceptron (MLP) serving as the adversary~\cite{Zhang2018, weerts2023fairlearn}.
The predictor model was trained to accurately detect FOG episodes while the adversary network tried to predict the protected attribute from the predictor's learned representations~\cite{Zhang2018}.
To learns attribute-agnostic features, the training minimizes the loss defined as $\nabla_WL_P - proj_{\nabla_WL_A}\nabla_WL_P - \alpha\nabla_WL_A$, where $L_P$ is the classification loss for the predictor model, $L_A$ is the adversary loss for the adversary model, and $\alpha$ is a hyperparameter that controls the strength with which the fairness constraint is enforced~\cite{Zhang2018}.
We experimented with two approaches: single attribute debiasing (training separate adversaries for each protected attribute) and a multi-head debiasing approach that implemented the adversary as a multi-head MLP to debias the protected attributes (FOG phenotype, age, sex, and disease duration) concurrently~\cite{Zhang2018}.
We hypothesized that multi-head adversarial debiasing would yield a fairer model across all attributes by simultaneously modeling their interplay.

\input{tables/transfer_learning_pair}

To address dataset homogeneity, we experimented with transfer learning across multi-site FOG datasets. 
We hypothesized that diversifying training data would simultaneously improve fairness and model performance.
To the best of our knowledge, this represents the first application of transfer learning for bias mitigation in FOG detection.
Since the datasets contained readings from different sensor locations (e.g., lower back, ankle), transfer learning required matching both sensor placements and sampling frequencies between the source and target datasets~\cite{chakma2021activity, link2022wearable}. 
Sensor locations were paired to ensure comparable on-body locations: lower extremity sensors (ankle, thigh, shank) for Daphnet and De Souza transfers, and lower-back sensors for multi-source combinations involving DeFOG (DeFOG + Daphnet $\rightarrow$ tDCS FOG; tDCS FOG + Daphnet $\rightarrow$ DeFOG). 
\add{
\autoref{tab:leave-one-dataset-out} shows the training and testing dataset pairs by sensor location for leave-one-dataset-out transfer learning.
}
Sampling frequencies were harmonized by downsampling all datasets to the lowest frequency in each pairing (e.g., 64 Hz when Daphnet was included).
For Model A2-A3, the first two convolutional layers were frozen while the final two convolutional layers and the recurrent layer were trained on target data, following standard transfer learning practice~\cite{Du2019, Hoelzemann2020}. 
For Model A1, pre-trained models were retrained on target datasets~\cite{kwon2020imutube}.
\add{Sec.~\ref{sec:zero-shot} shows the zero-shot learning performance for FOG detection and its bias without transfer learning.}

\par As a fourth bias mitigation approach, we investigated whether transfer learning from foundation models (models pretrained on large-scale, diverse population data) could reduce bias in FOG detection.
We hypothesized that rich activity representations learned from broader populations would transfer more equitably across demographic groups than models trained solely on limited FOG datasets. 
We evaluated two foundation model approaches that differed in their pretraining strategies and architectures: Model B1 pretrained via self-supervised learning on 700,000 person-days of wrist-worn accelerometer data from 100,000 UK Biobank participants~\cite{doherty2017large, yuan2024self}, and Model B2, which uses a transformer architecture pretrained on accelerometer data from 29,307 National Health and Nutrition Examination Survey (NHANES) participants~\cite{centers2019nhanes, ruan2025foundationmodelswearablemovement}.
For Model B1, we selected the variant trained on 5-second windows to match our FOG detection window length (3-5 seconds), modified only the output layer for binary FOG classification, and fine-tuned all layers during training on the four FOG datasets~\cite{yuan2024self}. 
For Model B2, we hypothesized that its transformer architecture for modeling long-range temporal dependencies~\cite{vaswani2017attention}, combined with pretraining on demographically diverse NHANES data, would yield robust representations transferable to FOG detection. 
To adapt the multi-channel FOG sensor data to Model B2's single-channel input requirement, we selected one sensor location per dataset and computed the magnitude from tri-axial accelerometer channels~\cite{ruan2025foundationmodelswearablemovement}. 
We froze all pretrained backbone weights and fine-tuned only the final classification layer on each FOG dataset~\cite{Du2019, Hoelzemann2020}.

\subsection{Model Performance and Bias Evaluation}
\label{sec:evaluation}


\par All models were evaluated using macro F1-scores under user-independent cross-validation to account for substantial class imbalance and ensure generalization to unseen subjects~\cite{opitz2019macro, kwon2019handling}.
The macro F1-score treats each class equally by computing the F1-score for each class independently, then averaging:
\begin{equation*}
    F1 = \dfrac{2}{|c|}\sum_{c}\dfrac{prec_c \times recall_c}{prec_c + recall_c}
\end{equation*}
where $|c|$ is the number of FOG classes, and $prec_c$ and $recall_c$ are the precision and recall for each class respectively~\cite{opitz2019macro}.
In clinical terms, recall represents sensitivity (the proportion of true FOG episodes correctly detected), while precision represents positive predictive value (the proportion of detected episodes that are truly FOG).
DeFOG, tDCS FOG, and De Souza datasets were evaluated using 5-fold user-independent cross-validation, while Daphnet used 3-fold user-independent cross-validation. 
Fewer folds for Daphnet (10 subjects) ensured each test fold included representation from both groups within each protected attribute (e.g., both male and female subjects for the sex attribute) for fairness evaluation.
Accounting for randomness in data splits and model training, we ran 10 iterations of cross-validation and reported average F1-scores across all test folds with 95\% confidence intervals for statistical significance.

\input{tables/thresholds}
\par Following standard group fairness evaluation~\cite{Dwork2012, Hardt2016}, we assessed whether models achieved comparable performance across subject groups defined by protected attributes (FOG phenotype, sex, age, disease duration).
As shown in \autoref{fig:fairml_pipeline}, subjects in each dataset were stratified into two groups per protected attribute.
To assess fairness across FOG phenotypes, episodes were classified as tremulous or akinetic based on spectral analysis of accelerometer data~\cite{moore2013autonomous}. 
\add{As mentioned in Sec.~\ref{sec:dataset},}
episodes with total power in the freezing band (3-8 Hz) exceeding that in the locomotion band (0-3 Hz) were classified as tremulous; remaining episodes were classified as akinetic~\cite{moore2013autonomous}.
\add{For all datasets, \autoref{tab:subject_fog_type} shows the FOG phenotype distribution per subject, based on each labeled episode.} 
For sex, we used the self-reported labels provided in the datasets (i.e., male and female).
Continuous attributes (age and disease duration) were dichotomized at dataset-specific medians following standard biostatistics practice~\cite{Gong2023}, as shown in~\autoref{tab:thresholds}.

\par Fairness was assessed using standard Fair ML metrics: Demographic Parity Ratio (DPR; Eq.~\eqref{eq:dpr}) and Equalized Odds Ratio (EOR; Eq.~\eqref{eq:eor}) for most attributes~\cite{pmlr-v80-agarwal18a}, and Equality of Opportunity Difference (EOD; Eq.~\eqref{eq:eod}) specifically for FOG phenotype~\cite{Hardt2016}. 
DPR was computed for FOG phenotype, age, sex, and disease duration, while EOR was computed for age, sex, and disease duration only. 
For the FOG phenotype, EOD was used instead of EOR because false positive rates cannot be computed given that FOG phenotype annotation is conditional on FOG episodes being present~\cite{moore2013autonomous}.
Similar to the F1-score, we reported the average DPR, EOR, and EOD, with 95\% confidence intervals from 10 iterations of cross-validation, to assess the statistical significance of model bias and our bias mitigation approaches.

\par For rigorous statistical significance testing, we employed the Wilcoxon signed-rank test to assess whether bias mitigation strategies produced statistically significant improvements in model performance (F1-score) and fairness metrics (DPR, EOR, and EOD)~\cite{wilcoxon1945individual}.
This non-parametric test was selected because it handles paired samples without requiring normality assumptions and is known to perform reliably with small sample sizes~\cite{wilcoxon1945individual}.
Since we focused on evaluating improvements in performance and fairness, a one-sided test was performed with the null hypothesis that median differences were zero, and significance was determined at $\alpha = 0.05$~\cite{wilcoxon1945individual}.
 These comparisons evaluated model performance and fairness before and after applying bias mitigation methods (threshold optimization, adversarial debiasing, and multi-site transfer), as well as the effect of fine-tuning large-scale pretrained models (Models B1-B2) relative to the baseline model (Model A2), which had been shown in prior literature to be effective for FOG detection~\cite{kwon2019handling}.


\par In the absence of FOG or PD-specific fairness thresholds, we adopted the four-fifths rule (threshold = 0.8), an established threshold from the U.S. Equal Employment Opportunity Commission that is widely used in Fair ML~\cite{eeoc_adverse_impact}, where DPR = 1 and EOR = 1 represent perfect fairness.

%% file: figures/fair_ml_pipeline.tex
\begin{figure*}[t!]
\centering
\includegraphics[width=\linewidth]{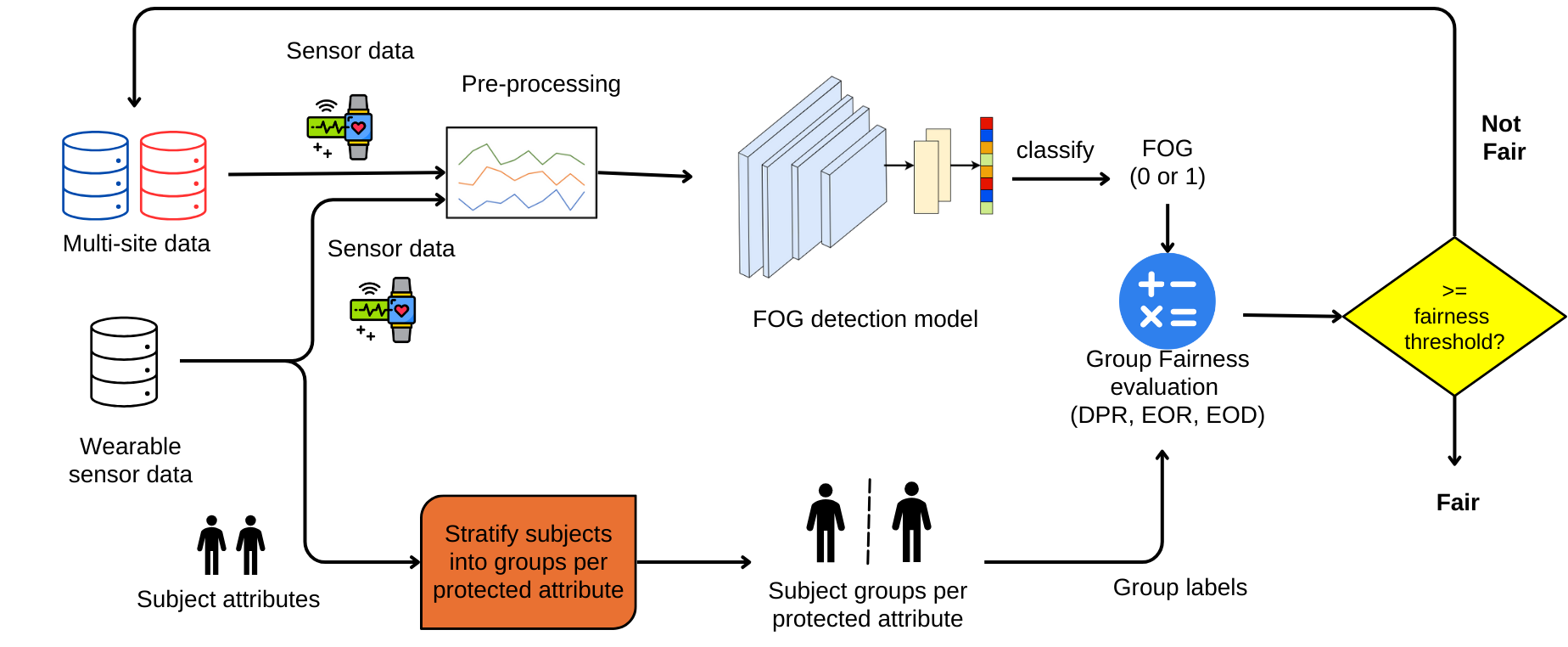}
\caption{Model bias evaluation pipeline for FOG detection. In this work, we analyze the group fairness of the model based on protected attributes in FOG and PD, including FOG phenotype, sex, age, and disease duration.}
\label{fig:fairml_pipeline}
\end{figure*}

%% file: tables/dataset_details_reviews.tex
\begin{table*}[t]
  \caption{
  Characteristics of datasets used in this study. 
  Numerical values are presented as mean $\pm$ standard deviation.
  N denotes the number of subjects, M refers to male, and F refers to female. 
  FOG type counts are the number of annotated timesteps of each phenotype, counted at the dataset's original sampling rate.}
  \label{tab:dataset_details_reviews}
  \begin{adjustbox}{width=\linewidth}
      \begin{tabular}{ccccccccccc}
        \toprule
         Dataset & N & Sex & \makecell{Age\\(years)} &\makecell{Disease\\Duration (years)} & \makecell{Frequency \\ (Hz)} & \makecell{Sensor\\locations} & \makecell{Label\\distribution} & \makecell{FOG type distribution\\ \add{proportion (samples)}} &\makecell{Average recording\\length per subject\\(minutes)} &\makecell{Total dataset\\recording length\\(hours)}\\
        \midrule
        Daphnet & 10 & \makecell{M=7\\F=3} & 66$\pm$5 & 13.7$\pm$9.7 & 64 & \makecell{Ankle,\\Lower back,\\Thigh} & \makecell{No FOG=0.90\\FOG=0.10} & \makecell{Tremulous=0.95 \add{(105,914)}\\Akinetic=0.05 \add{(4,871)}} & 29.6 $\pm$6.8 & 4.9\\
        
        De Souza & 35 & \makecell{M=19\\F=16} & 65$\pm$10 & 8.0$\pm$4.1 & 128 & \makecell{Shank} & \makecell{No FOG=0.80\\FOG= 0.20} & \makecell{Tremulous=0.97 \add{(200,321)}\\Akinetic=0.03 \add{(6,013)}} & 4.1$\pm$1.7 & 2.4\\
        
        DeFOG & 38 & \makecell{M=24\\F=14} & 67$\pm$9 & 9.5$\pm$5.5 & 100 & \makecell{Lower back} & \makecell{No FOG=0.95\\FOG=0.05} & \makecell{Tremulous=0.63 \add{(436,618)}\\Akinetic=0.37 \add{(249,887)}} & 59.3$\pm$31.7 & 15.0\\
        
        tDCS FOG & 62 & \makecell{M=50\\F=12} & 69$\pm$8 & 12.4$\pm$6.7 & 128 & \makecell{Lower back} & \makecell{No FOG=0.68\\FOG=0.32} & \makecell{Tremulous=0.86 \add{(1,888,654)}\\Akinetic=0.14 \add{(302,756)}} & 14.5$\pm$16.8 & 37.5\\
        
      \bottomrule 
    \end{tabular}
    \end{adjustbox}
\end{table*}

%% file: tables/transfer_learning_pair.tex
\begin{table*}[t]
\centering
\addBox{
\caption{
Training and testing dataset pairs used for leave-one-dataset-out cross-validation for transfer learning
}
\label{tab:leave-one-dataset-out}
\small
\centering
\begin{tabular}{ccc}
\toprule
Training Dataset & Test Dataset & Sensor locations \\
\midrule
DeFOG + Daphnet & DCS FOG & Lower back   \\
tDCS FOG + Daphnet  & DeFOG & Lower back   \\
Daphnet  & De Souza      & Ankle, Thigh, Shank   \\
De Souza  &  Daphnet     & Ankle, Thigh, Shank  \\
\bottomrule
\end{tabular}
}
\end{table*}

%% file: tables/thresholds.tex
\begin{table*}[t!]
  \centering
  \caption{Median values for continuous subject attributes}
   \begin{tabular}{ccccc}
    \toprule
     Attribute & Daphnet & De Souza & tDCS FOG & DeFOG\\
    \midrule
    Age (years) & 66 & 69 & 69 & 69 \\
    \makecell{Disease Duration\\(years)}  & 12.5 & 7 & 9 & 13 \\
  \bottomrule
\end{tabular}
\label{tab:thresholds}
\end{table*}

%% file: sections/results.tex
\section{Results}
In this section, we first report baseline model performance and fairness metrics, then evaluate the impact of the bias mitigation approaches.

\subsection{Model Performance and Bias without Bias Mitigation}

Model performance improved with architectural complexity across the multi-site datasets, as shown in \autoref{tab:results_before_mitigation} and \autoref{fig:premitigation_plot_boxplot}. 
Model A1 achieved an average F1-score of $0.578$, while the deep learning architectures demonstrated superior performance: Model A2 achieved $0.682$, and Model A3 achieved $0.688$. 
This progression reflected the increasing capacity of these architectures to capture temporal dependencies in wearable sensor data, with the attention-based Model A3 providing marginal gains over the convolutional-recurrent approach of Model A2.

Despite these performance gains, all models exhibited suboptimal fairness across clinical and demographic attributes, with no model achieving the 0.8 four-fifths rule threshold for any fairness metric (\autoref{tab:results_before_mitigation}).
FOG phenotype bias was particularly pronounced: DPR ranged from $0.435$ (Model A1) to $0.540$ (Model A3), and EOD ranged from $0.119$ to $0.233$, indicating systematic underperformance on underrepresented FOG phenotypes. 
Demographic attributes showed comparable fairness, with sex-based DPR values between $0.413$ and $0.514$, age-based DPR between $0.463$ and $0.486$, and disease duration-based DPR between $0.429$ and $0.509$. 
EOR values were consistently low, with all values below $0.42$ across protected attributes. 
Critically, increased model performance did not correspond to improved fairness, suggesting that standard training procedures introduced systematic biases that require targeted mitigation.

\input{figures/summary_plot_boxplot}

\subsection{Model Performance and Bias with Bias Mitigation}
\par Bias mitigation strategies yielded variable results, with most approaches failing to meet fairness thresholds across all attributes, including the FOG phenotype. 
Threshold optimization for Model A1 showed mixed outcomes with minimal performance impact (DPR: $-0.126$, EOR: $+0.063$, F1-score: $-0.011$), while adversarial debiasing applied independently to each attribute yielded mixed results for both Model A2 (DPR: $+0.005$, EOR: $-0.014$, F1-score: $-0.011$) and Model A3 (DPR: $-0.002$, EOR: $-0.021$, F1: $-0.017$) models. 
For FOG phenotype specifically, independent adversarial debiasing increased DPR by $+0.048$ (Model A2) and $+0.035$ (Model A3), though both remained well below fairness thresholds. 
Multi-head adversarial debiasing produced negligible overall changes in fairness metrics for both architectures (DPR: $+0.002$, EOR: $+0.007$).

\par Transfer learning approaches \del{substantially} outperformed other mitigation strategies in both fairness gains and performance improvements, although fairness thresholds were not met. 
Multi-site transfer learning increased F1-scores by $+0.024$ (Model A2) and $+0.017$ (Model A3), while improving DPR ($+0.037$) and EOR ($+0.045$) for both architectures. 
For the FOG phenotype, DPR improved by $+0.051$ (Model A2) and $+0.045$ (Model A3), respectively.
Transfer learning from generic activity representations (Model B1) improved F1-scores ($+0.015$) and DPR ($+0.027$) but showed minimal improvement for FOG phenotype (DPR: $+0.002$). 

\par Fine-tuning Model B2 produced the largest fairness gains, increasing DPR by $+0.159$ and EOR by $+0.131$ across all attributes, including FOG phenotype (DPR: $+0.157$), though this came at a substantial performance cost (F1-score: $-0.074$) and still fell short of fairness thresholds.
The improvements were consistent across demographic attributes, with DPR gains of $+0.156$ for sex, $+0.167$ for age (p < 0.05), and $+0.155$ for disease duration (p < 0.05), and EOR improvements ranging from $+0.123$ to $+0.146$. 
Despite representing the strongest fairness improvements among all tested approaches, most attributes remained below the four-fifths rule threshold across datasets.

\par Notably, the only experiment we conducted in which bias mitigation reduced bias to meet the four-fifths rule was Model B2 (foundation model feature transfer) on the tDCS FOG dataset.
This approach achieved DPR values exceeding 0.80 for all protected attributes (sex: $0.855$, age: $0.820$, disease duration: $0.850$, FOG phenotype: $0.939$), though at the cost of reduced F1-score ($0.558$ vs. $0.722$ baseline).

\input{tables/all_results_summarized}

Bias mitigation results are summarized in \autoref{tab:all_results_summarized}, with detailed performance (F1-score) and fairness metrics (DPR, EOR, and EOD) for all approaches shown in \autoref{tab:all_results}. 
Results for fine-tuning large-scale pretrained models for human activity are presented separately in \autoref{tab:all_results_foundation}.

%% file: figures/summary_plot_boxplot.tex
\begin{figure}[t]
\centering
\addBox{
\includegraphics[width=\linewidth]{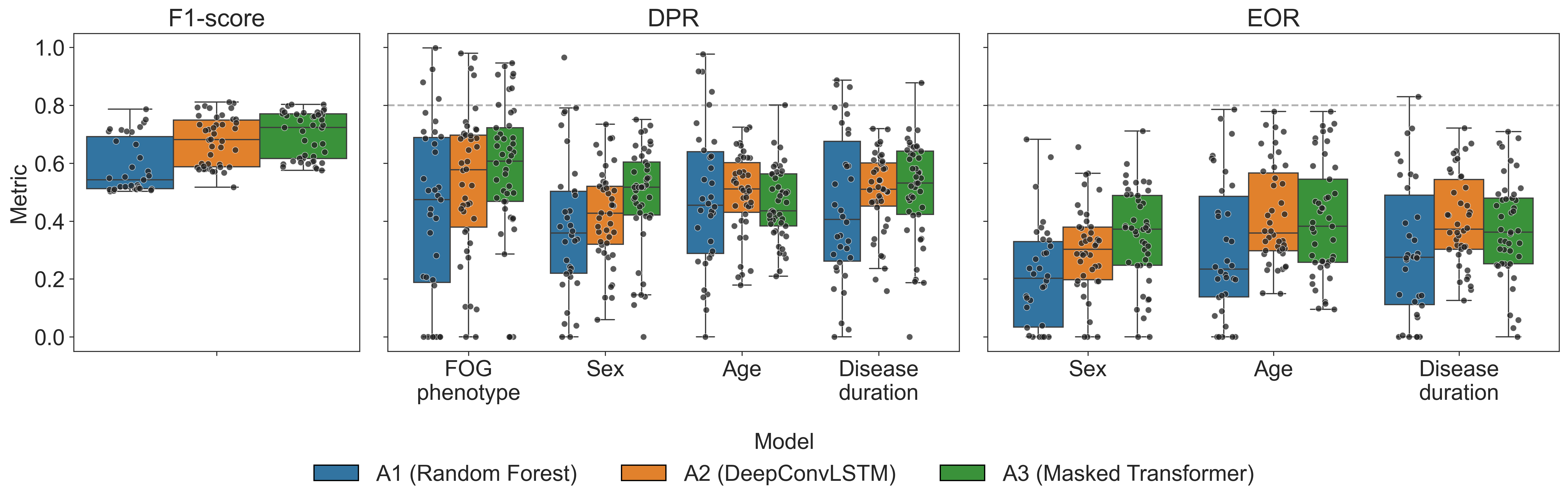}
\caption{
F1-score, Demographic Parity Ratio (DPR), and Equalized Odds Ratio (EOR) for each model across the multi-site datasets prior to applying bias mitigation approaches.
Dashed horizontal lines represent the fairness threshold (0.8 for DPR and EOR).
Each dot represents one iteration (a single seeded run of k-fold cross-validation) averaged across its folds.
}
\label{fig:premitigation_plot_boxplot}
}
\end{figure}

%% file: tables/all_results_summarized.tex
\begin{table*}[t]
\centering
\caption{Average change in performance (F1-score) and fairness (DPR, EOR, and EOD) across multi-site datasets after applying bias mitigation approaches (shown in the Bias Mitigation column). 
Statistical significance of changes was assessed using paired Wilcoxon signed-rank tests across datasets, comparing baseline and mitigation models. 
Each metric was first averaged within each dataset, and the Wilcoxon test was applied to these dataset-level averages to assess independence between pairs (Asterisks indicate significance levels: $^{*}$ p < 0.1\del{, $^{**}$ p < 0.05}).
}
\begin{adjustbox}{max width=\linewidth}
\small
\begin{tabular}{cc|ccc|cc|ccc|ccc}
\hline
\makecell{Bias\\Mitigation} & Model & $\Delta$F1-score & \makecell{$\Delta$DPR\\(all attributes)} & \makecell{$\Delta$EOR\\(Sex, Age,\\Disease\\duration)} & \makecell{$\Delta$DPR FOG\\phenotype} & \makecell{$\Delta$EOD FOG\\phenotype} & \makecell{$\Delta$DPR\\Sex} & \makecell{$\Delta$DPR\\Age} & \makecell{$\Delta$DPR\\Disease\\duration} & \makecell{$\Delta$EOR\\Sex} & \makecell{$\Delta$EOR\\Age} & \makecell{$\Delta$EOR\\Disease\\duration} \\
\hline
\makecell{Threshold\\optimizer} & A1 & -0.011 & -0.126 & +0.063 & +0.017 & +0.017 & -0.096 & -0.221 & -0.207 & +0.072 & +0.059 & +0.060 \\
\hline
\multirow{2}{*}{\makecell{Adversarial debiasing\\(Single attribute)}} & \makecell{A2} & -0.011 & +0.005 & -0.014 & +0.048 & -0.003 & +0.031 & -0.019 & -0.018 & +0.036 & -0.008 & -0.009 \\
    & \makecell{A3} & -0.017 & -0.002 & -0.021 & +0.035 & 0.000 & -0.060 & +0.017 & +0.007 & -0.067 & -0.001 & +0.003 \\ 
\hline
\multirow{2}{*}{\makecell{Adversarial debiasing\\(Multi-Head MLP)}} & \makecell{A2} & +0.001 & -0.019 & -0.011 & -0.001 & +0.015 & -0.002 & -0.033 & -0.029 & +0.004 & -0.002 & -0.033 \\
    & \makecell{A3} & -0.032 & +0.002 & +0.007 & +0.047 & -0.020 & -0.062 & +0.087 & -0.019 & -0.027 & +0.052 & -0.002 \\
\hline
\multirow{3}{*}{\makecell{Multi-site Transfer}} & A1 & -0.010 & -0.017 & -0.022 & -0.038 & -0.004 & +0.021 & -0.080 & +0.026 & -0.018 & -0.014 & -0.035 \\
 & \makecell{A2} & +0.024 & +0.037 & +0.045 & +0.051 & +0.028 & +0.038 & +0.043 & +0.016 & $+0.062^{*}$ & -0.041 & +0.033 \\
 & \makecell{A3} & +0.017 & +0.037 & +0.045 & +0.045 & +0.031 & -0.012 & $+0.045^{*}$ & +0.021 & +0.023 & +0.058 & +0.047 \\
\hline
\makecell{Generic Activity\\Feature Transfer} & B1 & +0.015 & +0.027 & -0.005 & +0.002 & +0.030 & +0.056 & +0.022 & +0.027 & +0.022 & -0.025 & -0.012 \\
\hline
\makecell{Foundation Model\\Feature Transfer} & B2 & -0.074 & +0.159 & +0.131 & +0.157 & -0.010 & +0.156 & $+0.167^{*}$ & $+0.155^{*}$ & +0.146 & +0.125 & +0.123 \\
\hline
\end{tabular}
\end{adjustbox}
\label{tab:all_results_summarized}
\end{table*}

%% file: sections/discussion.tex
\section{Discussion}

The main contribution of this work was the application of Fair ML frameworks to demonstrate systematic differences in FOG detection performance across FOG phenotypes and demographic strata.
We found that conventional bias mitigation techniques showed limited effectiveness, whereas multi-site transfer learning showed the most promise for reducing bias and improving overall performance. 
Our findings underscore systematic performance differences across FOG phenotypes, with the majority of models failing to meet fairness thresholds regardless of mitigation approach. 
Across majority datasets and experiments, DPR for FOG phenotypes remained below 0.8, indicating persistent bias favoring one phenotype over the other. 
This finding aligns with prior work by~\citet{yang2024automatic}, who reported that FOG detection models showed higher agreement with expert raters for tremulous FOG compared to akinetic FOG.

These findings suggest that post hoc algorithmic corrections alone, while helpful, may be inadequate when underlying data imbalance and phenotype heterogeneity are substantial. 
Critically, all bias mitigation approaches, including both conventional and transfer learning-based methods, failed to eliminate these phenotype-based disparities. 
Future efforts may require intentional sampling strategies that explicitly enrich underrepresented FOG phenotypes and demographic subgroups during data collection. 
This persistence suggests the performance differences arise from genuine pathophysiological differences in motor mechanisms between phenotypes~\cite{factor2025possible, zoetewei2025effects}. 
Such approaches could improve representation at the data level rather than relying solely on downstream mitigation. 
An alternative or complementary strategy may be to develop phenotype-specific or stratified models, in which separate models are trained for distinct FOG phenotypes. 
This may allow models to better capture phenotype-dependent signal characteristics that are otherwise obscured in pooled training.


\subsection{Model Performance and Bias without Bias Mitigation}


Deep learning architectures substantially outperformed shallow ML architectures for FOG detection across all datasets, with Model A2-A3 achieving macro-F1 scores that were $+0.104$ and $+0.110$ higher than Model A1, respectively.
This aligned with prior work by Kwon \textit{et al.} \cite{kwon2019handling}, which demonstrated superior FOG detection performance for deep learning architectures compared to shallow machine learning models by capturing temporal dependencies related to FOG.
The highest performance (macro F1-score of $0.778$) was achieved on the De Souza dataset, likely due to its simpler protocol, which focused on turning~\cite{boari2022}, in comparison to the more complex protocols in other datasets (e.g., walking, sitting, standing, opening doors).

Without bias mitigation, no protected attribute was found fair across the four datasets (DPR and EOR $\leq 0.8$). 
The tDCS FOG dataset had the highest DPR and EOR values across most protected attributes, with DPRs of $0.622$, $0.677$, and $0.715$ and EORs of $0.561$, $0.684$, and $0.624$ for sex, age, and disease duration, respectively.
This relatively better performance likely resulted from the tDCS FOG dataset's larger sample size (N=62), which provided greater representation compared to other datasets.
These findings demonstrated that adequate data representation is critical for bias mitigation.

More critically, the FOG phenotype showed bias (DPR < 0.8) across all datasets and models, due to significantly fewer akinetic than tremulous FOG episodes.
This imbalance reflected documented FOG phenotype distributions, as akinetic FOG occurs less frequently than tremulous FOG~\cite{zoetewei2025effects}, requiring more specific triggers (e.g., gait initiation) versus the more readily occurring tremulous episodes during general motion.

\subsection{Model Performance and Bias with Bias Mitigation}

\par The conventional bias mitigation approaches (i.e., threshold optimization, adversarial debiasing) yielded expected performance trade-offs, showing modest decreases in F1 score across models ($-0.032$ to $-0.010$). 
These reductions align with established findings in Fair ML literature: fairness constraints force models away from performance-optimal solutions by either imposing group-dependent decision thresholds (threshold optimization) or suppressing predictive information correlated with protected attributes (adversarial debiasing)~\cite{Hardt2016, Zhang2018, pmlr-v81-menon18a}


\par In contrast, two transfer learning approaches improved model performance while mitigating bias, with multi-site transfer yielding an average increase in F1 score of $+0.020$ and generic activity feature transfer increasing scores by $+0.015$.
Multi-site transfer leveraged the diversity across FOG datasets, exposing models to varied patient populations, sensor configurations, and clinical protocols during pretraining. 
This cross-dataset exposure enabled models to learn robust FOG representations that generalize beyond site-specific characteristics, reducing overfitting to the demographic and clinical distributions of any single dataset.
The increase in F1-score with Model B1 ($+0.015$) was consistent with the work by~\citet{yuan2024self}, who reported that their large generic activity recognition model was useful even in downstream health applications.

\par Despite testing multiple bias mitigation approaches, none met the fairness threshold, with only transfer learning methods showing meaningful fairness improvements.
Conventional bias mitigation techniques showed limited effectiveness, both producing mixed results across fairness metrics: threshold optimization with Model A1 ($-0.126$ DPR, $+0.063$ EOR), and single-attribute adversarial debiasing with Models A2-A3 ($+0.002$ DPR, $-0.018$ EOR).
In contrast, transfer learning approaches leveraged diverse, large-scale datasets to improve both fairness and performance, with multi-site transfer with Model A2-A3 ($+0.037$ DPR, $+0.045$ EOR) and generic activity feature transfer in Model B1 ($+0.027$ DPR) demonstrating that richer training data reduced bias.
Only one mitigation approach met the fairness threshold: transfer learning from Model B2 on one dataset (tDCS FOG), but at a substantial performance cost (a $-0.164$ decrease in F1-score).
Model B2's pretraining on the demographically diverse NHANES dataset likely enabled learning of demographic-invariant features, improving fairness but suppressing clinically relevant features specific to FOG detection.

\subsection{Relation to Current State of the Art}

Our models represent the state of the art in FOG detection when compared with successful solutions from the Kaggle Freezing of Gait detection challenge, which introduced the DeFOG and tDCS FOG datasets used in this study~\cite{salomon2024contest, kaggleFOG}. 
Model A3 (Masked Transformer) with multi-site transfer, pretrained on DeFOG and Daphnet and finetuned on tDCS FOG, achieved performance comparable to the competition's first-place solution (Model A3 F1-score = $0.762$ versus first-place F1-score = $0.790$)~\cite{salomon2024contest}.
Both approaches relied on transformer architectures that model temporal dependencies in FOG episodes~\cite{vaswani2017attention}. 
We evaluated our models using cross-validation on the public training data from the challenge rather than on the private test set used to rank competition submissions, limiting direct comparison with leaderboard results.
Additionally, the challenge leaderboard scored models using both F1-score and Mean Average Precision across FOG event classes (e.g., turn, start hesitation, walking)~\cite{salomon2024contest}, whereas we evaluated binary FOG detection using F1-score alone.
Nevertheless, since our models achieved state-of-the-art performance, the fairness analysis framework presented here could be applied to evaluate bias in competition-winning solutions.

\subsection{Limitations and Future Work}

While transfer learning approaches showed promise for improving both performance and fairness, several limitations warrant consideration.

\add{
Our findings provide encouraging evidence that transfer learning across diverse motor tasks and recording conditions can improve model performance across strata. At the same time, the available data impose important limits on the conclusions that can be drawn. The datasets used here differed substantially in their populations, tasks, acquisition protocols, and recording conditions (\autoref{tab:dataset_details_reviews} and \autoref{tab:dataset_protocol_details}), precluding a systematic ablation analysis of the specific contribution of transfer learning. A rigorous evaluation will require larger datasets designed to vary these factors in a controlled manner. 
Clinical attributes that may confound FOG manifestation were documented inconsistently across datasets (Table~\ref{tab:metadata}). 
Cognitive status was reported in only one dataset (De Souza), disease severity through non-comparable instruments (Hoehn and Yahr in Daphnet and De Souza, UPDRS-III in DeFOG and tDCS FOG), and FOG severity was absent in Daphnet.
These attributes could therefore not be controlled for consistently across sites.
Phenotype-specific disparities may partly reflect medication state, disease severity, cognitive status, or FOG severity rather than phenotype alone.
Future work may address this by collecting harmonized clinical metadata prospectively across sites to control for these confounds.
Despite these limitations, this study establishes an initial proof of concept and motivates more systematic investigation of transfer-learning strategies in future work.
}

\add{One major limitation is that, because expert annotation under contemporary criteria was not feasible within the available resources, we derived putative FOG phenotypes using an established spectral algorithm~\cite{moore2013autonomous}. This approach raises two concerns. First, although agreement with the available “Kinetic” annotations was acceptable (79.6\% accuracy; \autoref{fig:phenotype_annotation_confusion_matrix}), the algorithm’s dichotomous classification of akinetic versus tremulous FOG has been superseded by more fine-grained schemes. In particular, the most recent ICFOG classification divides Kinetic episodes into kinetic-trembling and kinetic-non-trembling FOG, because episodes without clear trembling or akinesia may be inconsistently assigned under the binary scheme~\cite{gilat2026updated}. Second, circular labeling or data leakage is possible because the phenotype labels and FOG classifiers were derived from the same underlying sensor data. Although the classifiers did not explicitly use the spectral features specified by the Moore algorithm~\cite{moore2013autonomous}, related information could have been learned indirectly. We consider substantial leakage unlikely, but it cannot be excluded. Future studies using finer-grained, expert-annotated phenotypes are needed to address these limitations. Nevertheless, fairness criteria may help detect systematic differences in classifier performance across phenotype groups, even when the group labels are imperfect.
}

\add{
Although we evaluated three widely used fairness criteria, the measures most appropriate for clinical applications remain an active area of research. Demographic parity, in particular, may be unsuitable when the protected attribute is causally associated with the outcome, because FOG prevalence may legitimately differ by, for example, phenotype or disease severity – indeed some reports suggest that this may be the case~\cite{santos2020impact,zhang2021clinical,perez2014prevalence}. Equal opportunity and equalized odds may arguably have greater clinical utility, as they generally assess whether true FOG episodes are detected equally well across groups, while allowing for prevalence differences across groups. However, without concrete evidence of demographic imbalance  we cannot be certain of this. Despite ongoing uncertainty regarding the optimal criterion, applying multiple fairness measures remains important for identifying and addressing performance disparities across clinically relevant subpopulations.
}

\add{
The “four-fifths” rule used in this study is common across several domains, although its widespread use does not necessarily make it ideal for the present application. Its relevance stems largely from its codification in U.S. federal guidelines, most prominently for employment selection~\cite{eeoc_uniformguidelines_1607_4D}. Also known as the 80\% rule, it identifies potential adverse impact when the selection rate for a racial, ethnic, or sex-based group is less than four-fifths of the rate for the group with the highest selection rate. Technologies that perform systematically worse for particular groups may expose developers and users to regulatory and legal concerns, making the rule a practically relevant benchmark. Nevertheless, it is an important but incomplete criterion for algorithmic fairness. Other criteria, performance measures, or cutoff values may be more appropriate for machine-learning applications such as FOG identification.
}

\par Based on the data we used here, we can speculate on the future of the state of the art. 
In particular, the foundation models (Model B1 and B2) were not necessarily trained on subjects with movement disorders, creating a domain gap between the pretraining data and the clinical FOG detection task~\cite{wiggins2022opportunities}.
Foundation models trained specifically on movement disorders populations could better capture the motor symptom heterogeneity characteristics of FOG and PD, potentially improving both fairness and performance. 
Such disease-specific pretraining may be essential for achieving consistent FOG detection across diverse patient populations with different phenotypes and demographics.

\par This study evaluated fairness with respect to individual attributes (FOG phenotype, sex, age, disease duration) but did not examine intersectional biases arising from combinations of protected attributes.
However, clinical evidence demonstrates that protected attributes interact to influence PD and FOG presentation, such as women developing PD at later ages than men~\cite{macht2007predictors,fullard2018sex}.
Future work may employ intersectional fairness frameworks~\cite{wang2022towards,lett2023translating} to evaluate model performance across multi-attribute patient groups and address biases.

\par We used dataset-specific optimal temporal window sizes from prior literature rather than a unified optimal window, which may have limited cross-dataset methodological consistency~\cite{daphnet_paper, reches2020using}.
Particularly, different window sizes may capture FOG phenotypes differently, with shorter windows potentially favoring brief akinetic episodes and longer windows favoring sustained tremulous freezing.
Future work may develop a data-driven approach to identify a single optimal temporal window size that generalizes across FOG phenotypes and datasets~\cite{li2018specialized}.

%% file: sections/conclusion.tex
\section{Conclusion}
FOG is a debilitating symptom of PD that is challenging to capture during clinical assessments, creating a need for wearable-based continuous monitoring. 
However, for such systems to expand specialist care equitably, models must perform consistently across  FOG phenotypes and demographics. 
This work provides the first systematic evaluation of bias and fairness in wearable-based FOG detection, evaluating state-of-the-art models and bias mitigation approaches (threshold optimization, adversarial debiasing, transfer learning) across multi-site datasets.
Our findings revealed performance differences across FOG phenotypes and demographic groups, with models frequently failing to meet the established four-fifths fairness threshold. 
While conventional mitigation approaches improved fairness, they did not do so consistently. 
Transfer learning using multi-site data and large-scale pretrained (foundation) models was more effective, though foundation models required performance trade-offs to achieve fairness thresholds. 
\add{
The results suggest transfer learning might partially reduce disparities, but not eliminate them fully.
The foundation model approach (Model B2) achieved the largest fairness improvements at the cost of a meaningful reduction in detection performance.
Most methods did not reliably meet the 0.8 fairness threshold across datasets and attributes, and where the threshold was met (Model B2 on tDCS FOG), F1 score performance was substantially reduced compared to baseline.
}
Critically, the FOG phenotype emerged as a significant source of bias, with akinetic episodes substantially underrepresented in datasets and associated with performance disparities, with direct implications for clinical care.
These findings show that achieving equitable FOG detection requires proactive considerations during data collection, including stratified sampling by phenotype and demographics, rather than relying solely on bias mitigation approaches.

%% file: sections/supplementary.tex
\section{Supplementary}
\renewcommand{\thetable}{S\arabic{table}}
\renewcommand{\thefigure}{S\arabic{figure}}
\setcounter{table}{0}
\setcounter{figure}{0}


\add{\subsection{Statistical Testing Framework and Details}}

\add{
\autoref{tab:stat_datasetlevel} shows the dataset-level Wilcoxon signed-rank tests supporting \autoref{tab:all_results_summarized}.
To assess whether each mitigation approach affected fairness and detection performance, we ran one-sided paired Wilcoxon signed-rank tests, with the dataset as the unit of analysis. 
Baseline and mitigated values from the same dataset formed each pair,
and a separate test was run for each (model, mitigation approach, fairness metric, sensitive attribute) combination for fairness and each (model, mitigation approach) combination for detection performance. 
Within each dataset, we first computed the mean value of the metric across all repeated runs of cross-validation under both the baseline and the mitigation approach, so that each of the four datasets (De Souza, Daphnet, DeFOG, tDCS FOG) contributes exactly one paired observation, yielding at most four matched pairs per test. 
Because DPR and EOR approach their ideal value of 1 with increasing fairness while EOD approaches its ideal value of 0, we tested the one-sided alternative that mitigation moves each
fairness metric toward its respective ideal. 
For F1-score, where larger values indicate better detection performance, we tested the one-sided alternative that the mitigated configuration attains a higher value. 
With four datasets, the exact signed-rank distribution has a minimum attainable one-sided $p$ of $2^{-4}=0.0625$, so $\alpha=0.05$ cannot be reached at this unit regardless of the size or the consistency of the underlying effect. 
We adopted a significance level of $\alpha=0.10$ and report these results as directionally consistent. 
}

\add{
Table~\ref{tab:resp_overall} shows the statistical tests supporting DPR, EOR, and F1 score values reported in the abstract were computed after testing whether multi-site transfer improved fairness and detection performance relative to the baseline, using
one-sided paired Wilcoxon signed-rank tests with baseline and mitigated values from the same dataset forming each pair, as described in Sec.~\ref{sec:evaluation}. 
These values summarize the two deep architectures (DeepConvLSTM and MaskedHAR).
For fairness, the unit of analysis was a (model, dataset, sensitive attribute) triple.
Four attributes for DPR (sex, age, disease duration, and FOG phenotype) and three for EOR, which is undefined for the phenotype grouping.
Each triple was treated as an independent sample. 
Attributes were retained as separate samples rather than averaged because bias in these models is attribute-specific. 
A mitigation approach may improve parity across sex while leaving phenotype-related disparity unchanged, and averaging over attributes would collapse precisely the distinction the analysis is intended to resolve.
Within each unit, we averaged the fairness metric or F1-score across repeated runs of cross-validation, yielding one matched pair per unit per configuration. 
Because DPR and EOR approach their ideal value of 1 with increasing fairness, and a higher F1-score indicates better detection performance, we tested the one-sided alternative that the transfer learning configuration attains higher values for each metric.
}

\add{\subsection{Agreement between IMU-derived and clinician-annotated FOG phenotype} 
We constructed the confusion matrix to illustrate the performance of the algorithmically derived FOG phenotype (our approach described in Sec.~\ref{sec:dataset}) versus the clinician-provided annotations in the tDCS FOG and DeFOG dataset documentation~\cite{kaggleFOG}. 
We sum per-recording true negative, false positive, false negative, and true positive sample counts across all 502 evaluated recordings (2,877,915 aligned timesteps) from tDCS FOG and DeFOG, so each sample contributes equally (\autoref{fig:phenotype_annotation_confusion_matrix}). 
Rows correspond to the clinician annotation and columns to the IMU-extracted label.
Cells report raw counts with row-wise percentages in parentheses, and row labels report the number and proportion of samples assigned to each clinician class. 
Overall accuracy was 79.6\% and balanced accuracy 64.2\%. 
Clinician annotations were highly imbalanced toward the Kinetic class (89.1\% of pooled samples), so a classifier assigning every sample to the majority class would exceed the observed accuracy. 
Balanced accuracy is therefore reported alongside accuracy and exceeded the 50\% chance level. 
Pooled counts are weighted towards tDCS FOG, which contributed 76.1\% of samples and showed higher agreement than defog (accuracy 86.9\% versus 56.3\%).
}

\add{
\subsection{Zero-shot model performance and fairness across datasets}
\label{sec:zero-shot}
}

\add{
To assess whether model performance and fairness generalize across datasets without fine-tuning on the target dataset, we evaluated the deep learning-based models (Models A2-A3) in a zero-shot setting. 
This mirrored potential deployment scenarios, where a FOG detection model would be applied to a new cohort without labeled data (zero-shot learning). 
Given that multi-site transfer learning improved FOG detection performance on several datasets (\autoref{tab:all_results}), we investigated whether the pretrained model learned a generalizable, unbiased representation applicable across subjects and study protocols.
We applied each pretrained model to the target dataset without fine-tuning, using the same leave-one-dataset cross-validation procedure and random seeds, similarly to the multi-site transfer experiments (\autoref{tab:leave-one-dataset-out}), and reported F1-score, DPR, EOR, and EOD across sex, age, disease duration, and FOG phenotype, with means and 95\% confidence intervals in \autoref{tab:zero_shot}.
}

\add{
Zero-shot transfer produced F1-scores above 0.5 on only one of four target datasets. 
On DeFOG, with tDCS FOG and Daphnet as sources, Model A2 reached an F1-score of $0.582$ and Model A3 reached $0.574$, with DPR ranging from $0.07$ to $0.46$ and EOR ranging from $0.15$ to $0.29$ across both models and all attributes. 
On the remaining three datasets, F1-scores ranged from $0.095$ to $0.450$ across both models. 
DPR and EOR were near zero on De Souza and tDCS FOG, and near 1 for Model A3 on Daphnet.
For fairness, near-zero DPR and EOR on De Souza and tDCS FOG reflected the absence of positive predictions rather than a disparity between groups, and near-unity DPR on Daphnet reflected uniform positive predictions across groups rather than equitable behavior. 
Equal prediction rates across demographic groups can arise when a model predicts almost exclusively one class, rather than reflecting equitable detection. 
}

\add{
Our evaluation showed that zero-shot generalizability was largely dataset-dependent.
Differences in sensor placement, sampling rate, recording environment, and FOG-provoking protocol shifted the input distribution so that learned representations do not transfer across datasets in a zero-shot setting.
DeFOG was a partial exception, where transfer succeeded with tDCS FOG and Daphnet as sources but failed in the reverse direction. 
This suggests that the asymmetry in zero-shot generalizability may reflect differences in source dataset size relative to the target (tDCS FOG has 62 subjects and 37.5 hours of recording, whereas DeFOG has 38 subjects and 15.0 hours). 
However, these feasibility experiments alone cannot determine whether sensor similarity, dataset size, or both drive the asymmetry.
These results have implications for both deployment and fairness evaluation. 
Deploying a FOG detection model to a new cohort without labeled target data risks failure and requires fine-tuning on target datasets.
}


\input{supplementary_tables/table_three_test_statistics}
\input{supplementary_tables/table_overall}
\input{figures/confusion_matrix}
\input{tables/all_results_before_mitigation}
\input{tables/all_results}
\input{tables/all_results_foundation}
\input{tables/dataset_protocol_details}

\input{tables/per_dataset_metadata_details}
\input{supplementary_tables/zero_shot_transfer}
\clearpage
\input{tables/subject_per_fog_type}

%% file: supplementary_tables/table_three_test_statistics.tex
\begin{table*}[t]
\addBox{
\centering
\caption{
Dataset-level Wilcoxon signed-rank tests supporting \autoref{tab:all_results_summarized}, one entry per (bias mitigation, model, metric) combination.
Each test compares baseline and mitigated dataset-level means across the four datasets ($n=4$ throughout).
Each cell gives the effect $\Delta$, the signed-rank statistic $W$, and the one-sided $p$-value.
DPR, EOR, and F1-score were tested against the alternative hypothesis that mitigation increases them. 
EOD was tested against the alternative hypothesis that mitigation decreases it.
The smallest attainable one-sided $p$ at $n=4$ is $2^{-4}=0.0625$, so no test at this unit can reach $p\leq0.05$.
Bold marks the five tests reaching $\alpha=0.10$.
Signed-rank tests report $W$ and $n$.
}
\label{tab:stat_datasetlevel}
\begin{adjustbox}{max width=\linewidth}
\footnotesize
\renewcommand{\arraystretch}{1.05}
\begin{tabular}{llccccccccc}
\hline
 &  &  & \multicolumn{4}{c}{DPR} & \multicolumn{3}{c}{EOR} & \\
\makecell{Bias\\Mitigation} & Model & F1 score & Sex & Age & \makecell{Disease\\duration} & \makecell{FOG\\phenotype} & Sex & Age & \makecell{Disease\\duration} & \makecell{EOD FOG\\phenotype} \\
\hline\hline
\makecell[l]{Threshold\\optimizer} & A1
 & \makecell{$\Delta{=}-0.011$\\$W{=}1$\\$p{=}1.000$} & \makecell{$\Delta{=}-0.096$\\$W{=}0$\\$p{=}0.500$} & \makecell{$\Delta{=}-0.221$\\$W{=}0$\\$p{=}0.500$} & \makecell{$\Delta{=}-0.207$\\$W{=}0$\\$p{=}0.500$} & \makecell{$\Delta{=}+0.017$\\$W{=}1$\\$p{=}1.000$}
 & \makecell{$\Delta{=}+0.072$\\$W{=}0$\\$p{=}0.500$} & \makecell{$\Delta{=}+0.059$\\$W{=}0$\\$p{=}0.500$} & \makecell{$\Delta{=}+0.060$\\$W{=}0$\\$p{=}0.500$} & \makecell{$\Delta{=}+0.017$\\$W{=}0$\\$p{=}1.000$} \\
\hline
\makecell[l]{Adversarial debiasing\\(Single attribute)} & A2
 & \makecell{$\Delta{=}-0.011$\\$W{=}3$\\$p{=}0.312$} & \makecell{$\Delta{=}+0.031$\\$W{=}4$\\$p{=}0.438$} & \makecell{$\Delta{=}-0.019$\\$W{=}7$\\$p{=}0.812$} & \makecell{$\Delta{=}-0.018$\\$W{=}1$\\$p{=}0.125$} & \makecell{$\Delta{=}+0.048$\\$W{=}4$\\$p{=}0.438$}
 & \makecell{$\Delta{=}+0.036$\\$W{=}2$\\$p{=}0.188$} & \makecell{$\Delta{=}-0.008$\\$W{=}3$\\$p{=}0.312$} & \makecell{$\Delta{=}-0.009$\\$W{=}1$\\$p{=}0.125$} & \makecell{$\Delta{=}-0.003$\\$W{=}9$\\$p{=}0.125$} \\
\hline
\makecell[l]{Adversarial debiasing\\(Single attribute)} & A3
 & \makecell{$\Delta{=}-0.017$\\$W{=}10$\\$p{=}1.000$} & \makecell{$\Delta{=}-0.060$\\$W{=}10$\\$p{=}1.000$} & \makecell{$\Delta{=}+0.017$\\$W{=}6$\\$p{=}0.688$} & \makecell{$\Delta{=}+0.007$\\$W{=}2$\\$p{=}0.188$} & \makecell{$\Delta{=}+0.035$\\$W{=}3$\\$p{=}0.312$}
 & \makecell{$\Delta{=}-0.067$\\$W{=}10$\\$p{=}1.000$} & \makecell{$\Delta{=}-0.001$\\$W{=}7$\\$p{=}0.812$} & \makecell{$\Delta{=}+0.003$\\$W{=}9$\\$p{=}0.938$} & \makecell{$\Delta{=}\phantom{+}0.000$\\$W{=}8$\\$p{=}0.188$} \\
\hline
\makecell[l]{Adversarial debiasing\\(Multi-Head MLP)} & A2
 & \makecell{$\Delta{=}+0.001$\\$W{=}3$\\$p{=}0.312$} & \makecell{$\Delta{=}-0.002$\\$W{=}4$\\$p{=}0.438$} & \makecell{$\Delta{=}-0.033$\\$W{=}5$\\$p{=}0.562$} & \makecell{$\Delta{=}-0.029$\\$W{=}5$\\$p{=}0.562$} & \makecell{$\Delta{=}-0.001$\\$W{=}4$\\$p{=}0.438$}
 & \makecell{$\Delta{=}+0.004$\\$W{=}2$\\$p{=}0.188$} & \makecell{$\Delta{=}-0.002$\\$W{=}4$\\$p{=}0.438$} & \makecell{$\Delta{=}-0.033$\\$W{=}5$\\$p{=}0.562$} & \makecell{$\Delta{=}+0.015$\\$W{=}9$\\$p{=}0.125$} \\
\hline
\makecell[l]{Adversarial debiasing\\(Multi-Head MLP)} & A3
 & \makecell{$\Delta{=}-0.032$\\$W{=}10$\\$p{=}1.000$} & \makecell{$\Delta{=}-0.062$\\$W{=}10$\\$p{=}1.000$} & \makecell{$\Delta{=}+0.087$\\$W{=}1$\\$p{=}0.125$} & \makecell{$\Delta{=}-0.019$\\$W{=}5$\\$p{=}0.562$} & \makecell{$\Delta{=}+0.047$\\$W{=}2$\\$p{=}0.188$}
 & \makecell{$\Delta{=}-0.027$\\$W{=}5$\\$p{=}0.562$} & \makecell{$\Delta{=}+0.052$\\$W{=}3$\\$p{=}0.312$} & \makecell{$\Delta{=}-0.002$\\$W{=}5$\\$p{=}0.562$} & \makecell{$\Delta{=}-0.020$\\$W{=}9$\\$p{=}0.125$} \\
\hline
\makecell[l]{Multi-site\\Transfer} & A1
 & \makecell{$\Delta{=}-0.010$\\$W{=}5$\\$p{=}0.875$} & \makecell{$\Delta{=}+0.021$\\$W{=}3$\\$p{=}0.625$} & \makecell{$\Delta{=}-0.080$\\$W{=}4$\\$p{=}0.750$} & \makecell{$\Delta{=}+0.026$\\$W{=}2$\\$p{=}0.375$} & \makecell{$\Delta{=}-0.038$\\$W{=}6$\\$p{=}1.000$}
 & \makecell{$\Delta{=}-0.018$\\$W{=}3$\\$p{=}0.625$} & \makecell{$\Delta{=}-0.014$\\$W{=}3$\\$p{=}0.625$} & \makecell{$\Delta{=}-0.035$\\$W{=}2$\\$p{=}0.375$} & \makecell{$\Delta{=}-0.004$\\$W{=}1$\\$p{=}0.875$} \\
\hline
\makecell[l]{Multi-site\\Transfer} & A2
 & \makecell{$\Delta{=}+0.024$\\$W{=}1$\\$p{=}0.125$} & \makecell{$\Delta{=}+0.038$\\$W{=}3$\\$p{=}0.312$} & \makecell{$\Delta{=}+0.043$\\$W{=}3$\\$p{=}0.312$} & \makecell{$\Delta{=}+0.016$\\$W{=}4$\\$p{=}0.438$} & \makecell{$\Delta{=}+0.051$\\$W{=}1$\\$p{=}0.125$}
 & \makecell{$\Delta{=}+0.062$\\$W{=}0$\\$p{=}\mathbf{0.062}$} & \makecell{$\Delta{=}-0.041$\\$W{=}2$\\$p{=}0.188$} & \makecell{$\Delta{=}+0.033$\\$W{=}2$\\$p{=}0.188$} & \makecell{$\Delta{=}+0.028$\\$W{=}6$\\$p{=}0.438$} \\
\hline
\makecell[l]{Multi-site\\Transfer} & A3
 & \makecell{$\Delta{=}+0.017$\\$W{=}0$\\$p{=}\mathbf{0.062}$} & \makecell{$\Delta{=}-0.012$\\$W{=}7$\\$p{=}0.812$} & \makecell{$\Delta{=}+0.045$\\$W{=}0$\\$p{=}\mathbf{0.062}$} & \makecell{$\Delta{=}+0.021$\\$W{=}3$\\$p{=}0.312$} & \makecell{$\Delta{=}+0.045$\\$W{=}1$\\$p{=}0.125$}
 & \makecell{$\Delta{=}+0.023$\\$W{=}4$\\$p{=}0.438$} & \makecell{$\Delta{=}+0.058$\\$W{=}4$\\$p{=}0.438$} & \makecell{$\Delta{=}+0.047$\\$W{=}3$\\$p{=}0.312$} & \makecell{$\Delta{=}+0.031$\\$W{=}3$\\$p{=}0.812$} \\
\hline
\makecell[l]{Generic Activity\\Feature Transfer} & B1
 & \makecell{$\Delta{=}+0.015$\\$W{=}3$\\$p{=}0.312$} & \makecell{$\Delta{=}+0.056$\\$W{=}3$\\$p{=}0.312$} & \makecell{$\Delta{=}+0.022$\\$W{=}5$\\$p{=}0.562$} & \makecell{$\Delta{=}+0.027$\\$W{=}3$\\$p{=}0.312$} & \makecell{$\Delta{=}+0.002$\\$W{=}3$\\$p{=}0.312$}
 & \makecell{$\Delta{=}+0.022$\\$W{=}3$\\$p{=}0.312$} & \makecell{$\Delta{=}-0.025$\\$W{=}6$\\$p{=}0.688$} & \makecell{$\Delta{=}-0.012$\\$W{=}6$\\$p{=}0.688$} & \makecell{$\Delta{=}+0.030$\\$W{=}4$\\$p{=}0.688$} \\
\hline
\makecell[l]{Foundation Model\\Feature Transfer} & B2
 & \makecell{$\Delta{=}-0.074$\\$W{=}10$\\$p{=}1.000$} & \makecell{$\Delta{=}+0.156$\\$W{=}1$\\$p{=}0.125$} & \makecell{$\Delta{=}+0.167$\\$W{=}0$\\$p{=}\mathbf{0.062}$} & \makecell{$\Delta{=}+0.155$\\$W{=}0$\\$p{=}\mathbf{0.062}$} & \makecell{$\Delta{=}+0.157$\\$W{=}2$\\$p{=}0.188$}
 & \makecell{$\Delta{=}+0.146$\\$W{=}1$\\$p{=}0.125$} & \makecell{$\Delta{=}+0.125$\\$W{=}1$\\$p{=}0.125$} & \makecell{$\Delta{=}+0.123$\\$W{=}1$\\$p{=}0.125$} & \makecell{$\Delta{=}-0.010$\\$W{=}8$\\$p{=}0.188$} \\
\hline
\end{tabular}
\end{adjustbox}
}
\end{table*}

%% file: supplementary_tables/table_overall.tex
\begin{table*}[htbp]
\centering
\addBox{
\caption{
Tests underlying the values quoted in the abstract: multi-site transfer versus the no-mitigation baseline, pooled across the four datasets and the two deep architectures (A2, DeepConvLSTM; A3, MaskedHAR). 
Each protected attribute as a separate unit.
}
\label{tab:resp_overall}
\begin{adjustbox}{max width=\linewidth}
\small
\begin{tabular}{llcccc}
\toprule
Specification & Metric & $n$ (pairs) & \makecell{Mean\\$\Delta$} & $p$-value & Test statistic \\
\midrule
\multirow{3}{*}{\makecell[l]{Primary\\(deep models, A2 + A3)}}
 & F1-score & 8  & $+0.020$ & $0.027$  & Wilcoxon $W=4$   \\
 & DPR      & 32 & $+0.038$ & $0.0034$ & Wilcoxon $W=122$ \\
 & EOR      & 24 & $+0.045$ & $0.0024$ & Wilcoxon $W=54$  \\
\bottomrule
\end{tabular}
\end{adjustbox}
}
\end{table*}

%% file: figures/confusion_matrix.tex
\begin{figure*}[t!]
\centering
\addBox{
\includegraphics[width=0.7\linewidth]{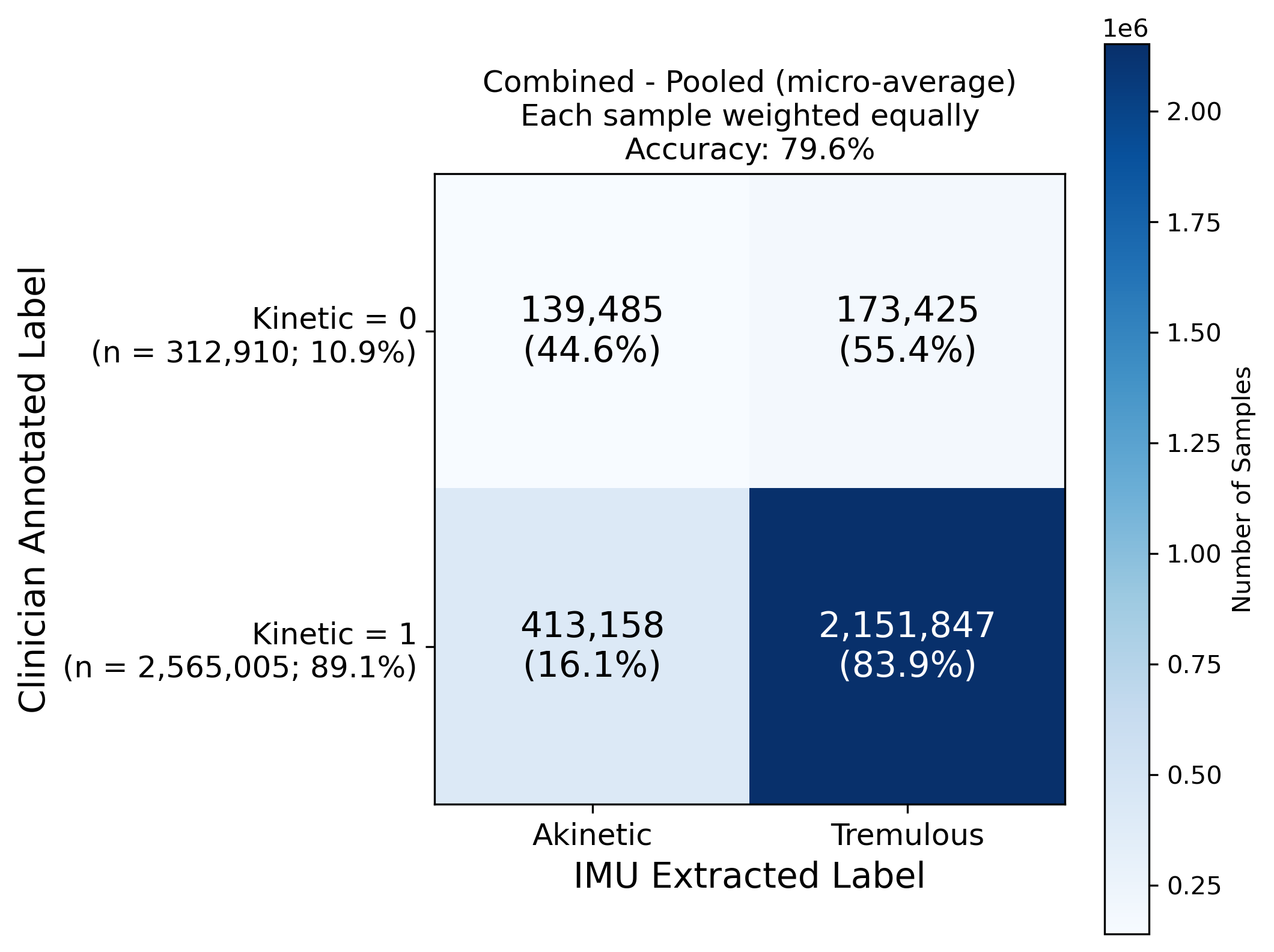}
\caption{
Confusion matrix between IMU-derived and clinician-annotated FOG phenotypes, pooled across the DeFOG and tDCS FOG (502 recordings,
2,877,915 samples).
}
\label{fig:phenotype_annotation_confusion_matrix}
}
\end{figure*}

%% file: tables/all_results_before_mitigation.tex
\begin{table*}[t]
\centering
\caption{Summary of the average performance (F1-score) and fairness (DPR, EOR, and EOD) across multi-site datasets before applying any bias mitigation approaches.}
\begin{adjustbox}{max width=\linewidth}
\small
\begin{tabular}{c|c|cc|ccc|ccc}
\hline
Model & F1-score & \makecell{DPR FOG\\phenotype} & \makecell{EOD FOG\\ phenotype} & \makecell{DPR\\Sex} & \makecell{DPR\\Age} & \makecell{DPR\\Disease\\duration} & \makecell{EOR\\Sex} & \makecell{EOR\\Age} & \makecell{EOR\\Disease\\duration} \\
\hline
A1 & 0.578 & 0.435 & 0.119 & 0.413 & 0.467 & 0.429 & 0.243 & 0.292 & 0.287 \\
\makecell{A2} & 0.682 & 0.527 & 0.226 & 0.438 & 0.486 & 0.500 & 0.323 & 0.416 & 0.408 \\
\makecell{A3} & 0.688 & 0.540 & 0.233 & 0.514 & 0.463 & 0.509 & 0.374 & 0.403 & 0.374 \\
\hline
\end{tabular}
\end{adjustbox}
\label{tab:results_before_mitigation}
\end{table*}

%% file: tables/all_results.tex
\begin{table*}[t!]
    \centering
    \caption{
    \rev{Model performance and fairness of the trained models for the  FOG detection task.
    Multi-site transfer occurs when models are pretrained on source datasets (in the Pretrain dataset column). 
    \textbf{Bold} text indicates the best value in each column for each dataset. 
    $^{*}$ indicates statistical significance ($p \leq 0.05$, Wilcoxon Signed Rank test~\cite{woolson2007wilcoxon}) comparing each bias mitigation method to its corresponding no-mitigation baseline ($\times$) within the same model and dataset. 
    }}
    \begin{adjustbox}{max width=\linewidth}
    \small
    \begin{tabular}{ccccc|cccc|cccc}
        \hline
          &    &   &   &   &  \multicolumn{4}{c|}{Demographic Parity Ratio (DPR)}  &  \multicolumn{3}{c}{Equalized Odds Ratio (EOR)} & \makecell{Equality of \\Opportunity\\Difference (EOD)}\\
        Dataset  &  Model  &  \makecell{Bias\\mitigation}  &  \makecell{Pretrain\\dataset}  &  F1-score  &  Sex  &  Age  &  \makecell{Disease\\duration}  &  \makecell{FOG phenotype}  & Sex  &  Age  &  \makecell{Disease\\duration}  &  \makecell{FOG phenotype}\\
        \hline\hline
        \multirow{8}{*}[-4em]{DeFOG}  &  \multirow{4}{*}[-2em]{A1}  &  $\times$  &  -  &  0.511 $\pm$ 0.008 & 0.332 $\pm$ 0.347 & 0.512 $\pm$ 0.311 & 0.279 $\pm$ 0.276 & \textbf{0.574 $\pm$ 0.212} & 0.141 $\pm$ 0.160 & 0.219 $\pm$ 0.205 & 0.203 $\pm$ 0.289 & \textbf{0.027 $\pm$ 0.019}\\
          &   &  \makecell{Threshold\\optimizer}  &  -  &  0.509 $\pm$ 0.004 & 0.288 $\pm$ 0.108 & 0.349 $\pm$ 0.114 & 0.374 $\pm$ 0.126 & 0.489 $\pm$ 0.124 & 0.256 $\pm$ 0.108 & 0.283 $\pm$ 0.124 & 0.216 $\pm$ 0.070 & 0.035 $\pm$ 0.039 \\
          &   &  \makecell{Multi-site\\ Transfer}  &  \makecell{tDCS FOG\\+ Daphnet}  &  0.511 $\pm$ 0.008 & 0.213 $\pm$ 0.239 & 0.260 $\pm$ 0.201 & 0.261 $\pm$ 0.199 & 0.524 $\pm$ 0.272 & 0.109 $\pm$ 0.187 & 0.096 $\pm$ 0.132 & 0.081 $\pm$ 0.115 & 0.027 $\pm$ 0.019 \\
         \cline{2-13}
          &  \multirow{4}{*}[-2em]{\makecell{A2}}  &  $\times$  &  -  &  0.582 $\pm$ 0.010  &  0.402 $\pm$ 0.068  &  0.528 $\pm$ 0.076  &  0.569 $\pm$ 0.077  &  0.418 $\pm$ 0.064  &  0.300 $\pm$ 0.054  &  0.320 $\pm$ 0.058  &  0.382 $\pm$ 0.064  &  0.183 $\pm$ 0.027 \\
          &   &  \makecell{Adversarial debiasing\\(Single attribute)}  &  -  &  0.575 $\pm$ 0.009  &  0.391 $\pm$ 0.090  &  0.406 $\pm$ 0.101  &  0.461 $\pm$ 0.086  & 0.422 $\pm$ 0.080  &  0.287 $\pm$ 0.075  &  0.270 $\pm$ 0.074  &  0.348 $\pm$ 0.089  & 0.152 $\pm$ 0.036 \\         
         &   &  \makecell{Adversarial debiasing\\(Multi-Head MLP)}  &  -  &  0.584 $\pm$ 0.017 &	0.482 $\pm$ 0.101 &	0.353 $\pm$ 0.088 & 0.428 $\pm$ 0.106 &	0.402 $\pm$ 0.099 &	0.346 $\pm$ 0.087 &	0.241 $\pm$ 0.069 &	0.280 $\pm$ 0.085 &	0.173 $\pm$ 0.042 \\  
          &   &  \makecell{Multi-site\\ Transfer}  &  \makecell{tDCS FOG\\+ Daphnet}  &  $\mathbf{{0.644 \pm 0.012}^{*}}$ & ${0.486 \pm 0.073}^{*}$ & \textbf{0.556 $\pm$ 0.068} & 0.562 $\pm$ 0.071 & ${0.494 \pm 0.058}^{*}$ & 0.356 $\pm$ 0.055 & $\mathbf{{0.429 \pm 0.069}^{*}}$ & $\mathbf{0.465 \pm 0.068}^{*}$ & 0.256 $\pm$ 0.035 \\
         \cline{2-13}
          &  \multirow{4}{*}[-2em]{\makecell{A3}}  &  $\times$  &  -  &   0.598 $\pm$ 0.014 & \textbf{0.620 $\pm$ 0.070} & 0.505 $\pm$ 0.078 & \textbf{0.618 $\pm$ 0.077} & 0.467 $\pm$ 0.068 & \textbf{0.426 $\pm$ 0.071} & 0.369 $\pm$ 0.071 & 0.393 $\pm$ 0.057 & 0.188 $\pm$ 0.031 \\
          &   &  \makecell{Adversarial debiasing\\(Single attribute)}  &  -  & 0.592 $\pm$ 0.010 &	0.553 $\pm$ 0.118 &	0.486 $\pm$ 0.112 &	0.572 $\pm$ 0.089 &	0.471 $\pm$ 0.108 &	0.289 $\pm$ 0.086 &	0.317 $\pm$ 0.097 &	0.367 $\pm$ 0.077 &	0.177 $\pm$ 0.051 \\
         &   &  \makecell{Adversarial debiasing\\(Multi-Head MLP)}  &  -  &  0.587 $\pm$ 0.011 & 0.464 $\pm$ 0.073 & 0.453 $\pm$ 0.072 & 0.523 $\pm$ 0.068 & 0.466 $\pm$ 0.095 & 0.334 $\pm$ 0.065 & 0.305 $\pm$ 0.062 & 0.347 $\pm$ 0.059 & 0.176 $\pm$ 0.038 \\
          &   &  \makecell{Multi-site\\ Transfer}  &  \makecell{tDCS FOG\\+ Daphnet}  &  0.631 $\pm$ 0.011 & 0.560 $\pm$ 0.050 & 0.504 $\pm$ 0.057 & 0.561 $\pm$ 0.053 & 0.546 $\pm$ 0.053 & 0.415 $\pm$ 0.049 & 0.383 $\pm$ 0.049 & 0.372 $\pm$ 0.052 & 0.221 $\pm$ 0.030\\
        \hline\hline
        \multirow{8}{*}[-4em]{tDCS FOG}  &  \multirow{4}{*}[-2em]{A1}  &  $\times$  &  -  &  0.515 $\pm$ 0.006 & 0.622 $\pm$ 0.144 & \textbf{0.677 $\pm$ 0.072} & \textbf{0.715 $\pm$ 0.071} & 0.610 $\pm$ 0.124 & 0.487 $\pm$ 0.180 & 0.586 $\pm$ 0.072 & 0.524 $\pm$ 0.099 & 0.066 $\pm$ 0.024 \\
          &   &  \makecell{Threshold\\optimizer}  &  -  &  0.519 $\pm$ 0.006 & 0.573 $\pm$ 0.147 & 0.189 $\pm$ 0.032 & 0.175 $\pm$ 0.033 & 0.702 $\pm$ 0.107 & 0.550 $\pm$ 0.188 & 0.690 $\pm$ 0.104 & 0.603 $\pm$ 0.095 & \textbf{0.066 $\pm$ 0.024} \\
          &    &  \makecell{Multi-site\\ Transfer}  &  \makecell{DeFOG\\+ Daphnet}  &  0.513 $\pm$ 0.007 & 0.539 $\pm$ 0.147 & 0.648 $\pm$ 0.084 & 0.711 $\pm$ 0.095 & 0.545 $\pm$ 0.126 & 0.365 $\pm$ 0.170 & 0.501 $\pm$ 0.080 & 0.496 $\pm$ 0.115 & 0.085 $\pm$ 0.033 \\
         \cline{2-13}
          &  \multirow{4}{*}[-2em]{\makecell{A2}}  &  $\times$  &  -  &  0.722 $\pm$ 0.019  &  0.614 $\pm$ 0.081  &  0.598 $\pm$ 0.064  &  0.591 $\pm$ 0.060  &  0.660 $\pm$ 0.060   &  0.543 $\pm$ 0.083  &  0.658 $\pm$ 0.054  &  0.624 $\pm$ 0.053  & 0.228 $\pm$ 0.040 \\
          &   &  \makecell{Adversarial debiasing\\(Single attribute)}  &  -  &  0.718 $\pm$ 0.013  &  0.654 $\pm$ 0.091  &  0.644 $\pm$ 0.077  &  0.605 $\pm$ 0.064  & $\mathbf{0.775 \pm 0.055}^{*}$  &  0.597 $\pm$ 0.097  &  0.665 $\pm$ 0.056  &  0.594 $\pm$ 0.053  & 0.118 $\pm$ 0.033 \\
         &   &  \makecell{Adversarial debiasing\\(Multi-Head MLP)}  &  -  &  0.710 $\pm$ 0.024 &	0.515 $\pm$ 0.073 &	0.590 $\pm$ 0.061 &	0.534 $\pm$ 0.060 &	0.671 $\pm$ 0.057 &	0.452 $\pm$ 0.073 &	0.622 $\pm$ 0.047 &	0.561 $\pm$ 0.055 &	0.203 $\pm$ 0.038 \\
          &   &  \makecell{Multi-site\\ Transfer}  &  \makecell{DeFOG\\+ Daphnet}  &  $\mathbf{0.752 \pm 0.022}^{*}$  &  0.658 $\pm$ 0.078  &  0.578 $\pm$ 0.054  &  0.626 $\pm$ 0.056  &  ${0.747 \pm 0.043}^{*}$  &  $\mathbf{{0.636 \pm 0.072}^{*}}$ &  0.704 $\pm$ 0.055  &  $\mathbf{{0.680 \pm 0.044}^{*}}$  &  0.214 $\pm$ 0.038 \\
        \cline{2-13} 
          &  \multirow{4}{*}[-2em]{\makecell{A3}}  &  $\times$  &  -  &  0.758 $\pm$ 0.018 & 0.622 $\pm$ 0.074 & 0.558 $\pm$ 0.064 & 0.564 $\pm$ 0.055 & 0.650 $\pm$ 0.045 & 0.561 $\pm$ 0.074 & 0.684 $\pm$ 0.050 & 0.593 $\pm$ 0.057 & 0.267 $\pm$ 0.040\\
          &   &  \makecell{Adversarial debiasing\\(Single attribute)}  &  -  &  0.738 $\pm$ 0.011 &	0.613 $\pm$ 0.109 &	0.572 $\pm$ 0.072 &	0.592 $\pm$ 0.059 & 0.741 $\pm$ 0.044 &	0.494 $\pm$ 0.108 &	0.655 $\pm$ 0.063 &	0.597 $\pm$ 0.065 &	0.179 $\pm$ 0.037\\
         &   &  \makecell{Adversarial debiasing\\(Multi-Head MLP)}  &  -  &  0.717 $\pm$ 0.021 & 0.627 $\pm$ 0.065 & 0.576 $\pm$ 0.048 & 0.602 $\pm$ 0.050 & 0.739 $\pm$ 0.063 & 0.536 $\pm$ 0.070 & 0.635 $\pm$ 0.046 & 0.614 $\pm$ 0.055 & 0.142 $\pm$ 0.034\\
          &   &  \makecell{Multi-site\\ Transfer}  &  \makecell{DeFOG\\+ Daphnet}  &  \textbf{0.762 $\pm$ 0.020} & $\textbf{0.654 $\pm$ 0.071}$ & 0.571 $\pm$ 0.061 & 0.611 $\pm$ 0.056 & 0.706 $\pm$ 0.057 & 0.599 $\pm$ 0.078 & \textbf{0.718 $\pm$ 0.050} & 0.647 $\pm$ 0.051 & 0.243 $\pm$ 0.047\\
        \hline\hline
        
        \multirow{8}{*}[-4em]{De Souza}  &  \multirow{4}{*}[-2em]{A1}  &  $\times$  &  -   &  0.724 $\pm$ 0.025&0.326 $\pm$ 0.080&0.340 $\pm$ 0.076&0.315 $\pm$ 0.072&0.488 $\pm$ 0.125&0.219 $\pm$ 0.069&0.224 $\pm$ 0.061&0.217 $\pm$ 0.075 & 0.315 $\pm$ 0.083 \\
         &   &  \makecell{Threshold\\optimizer}  &  -  &  0.705 $\pm$ 0.025&0.296 $\pm$ 0.078&0.345 $\pm$ 0.080&0.311 $\pm$ 0.086& 0.508 $\pm$ 0.132 & 0.264 $\pm$ 0.062&0.319 $\pm$ 0.070&0.317 $\pm$ 0.073 & 0.315 $\pm$ 0.083\\
          &   &  \makecell{Multi-site\\ Transfer}  &  Daphnet  & 0.724 $\pm$ 0.025&0.326 $\pm$ 0.080&0.340 $\pm$ 0.076&0.315 $\pm$ 0.072&0.488 $\pm$ 0.125&0.219 $\pm$ 0.069&0.224 $\pm$ 0.061&0.217 $\pm$ 0.075 & 0.315 $\pm$ 0.083 \\
         \cline{2-13}
          &  \multirow{4}{*}[-2em]{\makecell{A2}}  &  $\times$  &  -  &  \textbf{0.778 $\pm$ 0.033}  &  \textbf{0.464 $\pm$ 0.080}  &  0.423 $\pm$ 0.078  &  \textbf{0.481 $\pm$ 0.082}  &  \textbf{0.786 $\pm$ 0.078}  &  0.317 $\pm$ 0.060  &  0.362 $\pm$ 0.073  &  0.404 $\pm$ 0.075  &  \textbf{0.191 $\pm$ 0.070}\\
          &   &  \makecell{Adversarial debiasing \\(Single attribute)}  &  -  &  0.777 $\pm$ 0.022  &  0.455 $\pm$ 0.082  &  0.446 $\pm$ 0.080  &  0.472 $\pm$ 0.084  &  0.541 $\pm$ 0.112  &  0.328 $\pm$ 0.074  &  \textbf{0.391 $\pm$ 0.086}  &  \textbf{0.416 $\pm$ 0.082}  & 0.308 $\pm$ 0.082 \\
         &   &  \makecell{Adversarial debiasing \\(Multi-Head MLP)}  &  -  &  0.785 $\pm$ 0.027 &	0.411 $\pm$ 0.055 &	\textbf{0.460 $\pm$ 0.057} & 0.467 $\pm$ 0.059 & 0.653 $\pm$ 0.078 & 0.315 $\pm$ 0.046 & 0.409 $\pm$ 0.063 & 0.376 $\pm$ 0.053 & 0.258 $\pm$ 0.059 \\
         &   &  \makecell{Multi-site\\ Transfer}  &  Daphnet  &  0.774 $\pm$ 0.034 & 0.458 $\pm$ 0.083 & 0.399 $\pm$ 0.074 & 0.436 $\pm$ 0.074 & 0.757 $\pm$ 0.101 & \textbf{0.369 $\pm$ 0.072} & 0.283 $\pm$ 0.058 & 0.356 $\pm$ 0.065 & 0.208 $\pm$ 0.087\\
         \cline{2-13}
          &  \multirow{4}{*}[-2em]{\makecell{A3}}  &  $\times$  &  -  &  0.769 $\pm$ 0.030 & 0.410 $\pm$ 0.067 & 0.383 $\pm$ 0.055 & 0.465 $\pm$ 0.069 & 0.779 $\pm$ 0.066 & 0.307 $\pm$ 0.060 & 0.328 $\pm$ 0.057 & 0.322 $\pm$ 0.054 & 0.202 $\pm$ 0.057\\
          &   &  \makecell{Adversarial debiasing \\(Single attribute)}  &  -  &  0.768 $\pm$ 0.017 &	0.392 $\pm$ 0.086	&0.387 $\pm$ 0.065&	0.439 $\pm$ 0.057& 0.610 $\pm$ 0.113 &	0.316 $\pm$ 0.079	&0.353 $\pm$ 0.075	&0.320 $\pm$ 0.053	& 0.276 $\pm$ 0.084  \\
         &   &  \makecell{Adversarial debiasing \\(Multi-Head MLP)}  &  -  &  0.770 $\pm$ 0.038 & 0.412 $\pm$ 0.089 & ${0.417 \pm 0.081}^{*}$ & 0.444 $\pm$ 0.091 & 0.584 $\pm$ 0.084 & 0.300 $\pm$ 0.076 & 0.347 $\pm$ 0.089 & 0.308 $\pm$ 0.068 & 0.320 $\pm$ 0.069\\
         &   &  \makecell{Multi-site\\ Transfer}  &  Daphnet  &  0.758 $\pm$ 0.036 & 0.387 $\pm$ 0.087 & 0.376 $\pm$ 0.071 & 0.448 $\pm$ 0.080 & 0.769 $\pm$ 0.076 & 0.287 $\pm$ 0.077 & 0.253 $\pm$ 0.065 & 0.317 $\pm$ 0.070 & 0.213 $\pm$ 0.076 \\
        \hline\hline
        \multirow{8}{*}[-4em]{Daphnet}  &  \multirow{4}{*}[-2em]{A1}
         &  $\times$  &  -   &  0.560 $\pm$ 0.026	& 0.370 $\pm$ 0.244 &	0.341 $\pm$ 0.181 &	0.405 $\pm$ 0.181 &	0.068 $\pm$ 0.102 &	0.126 $\pm$ 0.154 &	0.139 $\pm$ 0.136 &	0.203 $\pm$ 0.149 &	0.124 $\pm$ 0.042 \\
         &   &  \makecell{Threshold\\optimizer}  &  -  &  0.533 $\pm$ 0.010 &	0.109 $\pm$ 0.060 &	0.101 $\pm$ 0.113 &	0.023 $\pm$ 0.013 & 0.129 $\pm$ 0.086 &	0.190 $\pm$ 0.130 &	0.112 $\pm$ 0.083	& 0.252 $\pm$ 0.108 & 0.214 $\pm$ 0.114 \\
         &   &  \makecell{Multi-site\\ Transfer}  &  De Souza  &  0.523 $\pm$ 0.014 & \textbf{0.655 $\pm$ 0.194}	& 0.303 $\pm$ 0.464 &	0.536 $\pm$ 0.283 &	0.032 $\pm$ 0.063 &	0.207 $\pm$ 0.278 &	0.293 $\pm$ 0.444 & 0.207 $\pm$ 0.278 &	0.214 $\pm$ 0.114\\
        \cline{2-13}
         &  \multirow{4}{*}[-2em]{\makecell{A2}}  &  $\times$  &  -  &  0.613 $\pm$ 0.037  &  0.330 $\pm$ 0.114  &  0.383 $\pm$ 0.101  &  0.401 $\pm$ 0.105  &  0.245 $\pm$ 0.138  &  0.185 $\pm$ 0.090  &  0.312 $\pm$ 0.107  &  0.285 $\pm$ 0.108  &  0.303 $\pm$ 0.094 \\
         &   &  \makecell{Adversarial debiasing\\(Single Attribute)}  &  -  &  0.602 $\pm$ 0.022  &  0.378 $\pm$ 0.165  &  0.368 $\pm$ 0.101  &  0.377 $\pm$ 0.131  & 0.180 $\pm$ 0.158  &  0.223 $\pm$ 0.137  &  0.306 $\pm$ 0.123  &  0.241 $\pm$ 0.155  & 0.317 $\pm$ 0.104 \\
         &   &  \makecell{Adversarial debiasing\\(Multi MLP Head)}  &  -  &  0.649 $\pm$ 0.025 &	0.334 $\pm$ 0.115 &	0.407 $\pm$ 0.115 &	0.442 $\pm$ 0.118 &	0.337 $\pm$ 0.109 &	0.194 $\pm$ 0.091 &	0.383 $\pm$ 0.114	& 0.284 $\pm$ 0.107  & 0.332 $\pm$ 0.059 \\
         &   &  \makecell{Multi-site\\ Transfer}  &  De Souza  &  0.651 $\pm$ 0.023 & 0.301 $\pm$ 0.124 & ${0.579 \pm 0.106}^{*}$ & 0.425 $\pm$ 0.116 & 0.316 $\pm$ 0.147 & 0.177 $\pm$ 0.105 & 0.409 $\pm$ 0.099 & 0.268 $\pm$ 0.109 & 0.337 $\pm$ 0.089 \\
        \cline{2-13}
         &  \multirow{4}{*}[-2em]{\makecell{A3}}  &  $\times$  &  -  &  0.626 $\pm$ 0.026 & 0.404 $\pm$ 0.161 & 0.405 $\pm$ 0.128 & 0.388 $\pm$ 0.145 & 0.262 $\pm$ 0.143 & 0.202 $\pm$ 0.114 & 0.229 $\pm$ 0.105 & 0.188 $\pm$ 0.103 & 0.276 $\pm$ 0.066\\
         &   &  \makecell{Adversarial debiasing\\(Single Attribute)}  &  -  & 0.586 $\pm$ 0.013 &	0.260 $\pm$ 0.139 &	0.473 $\pm$ 0.165&	0.460 $\pm$ 0.155	& \textbf{0.439 $\pm$ 0.149}	&0.129 $\pm$ 0.086&	0.281 $\pm$ 0.144	&0.224 $\pm$ 0.127	& \textbf{0.187 $\pm$ 0.070}\\
         &   &  \makecell{Adversarial debiasing\\(Multi MLP Head)}  &  -  &  0.578 $\pm$ 0.024 & 0.350 $\pm$ 0.152 & 0.417 $\pm$ 0.143 & 0.405 $\pm$ 0.140 & 0.385 $\pm$ 0.120 & 0.137 $\pm$ 0.093 & 0.256 $\pm$ 0.123 & 0.211 $\pm$ 0.111 & 0.214 $\pm$ 0.057\\
         &   &  \makecell{Multi-site\\ Transfer}  &  De Souza  &  \textbf{0.667 $\pm$ 0.026} & 0.409 $\pm$ 0.143 & \textbf{0.581 $\pm$ 0.089} & \textbf{0.499 $\pm$ 0.128} & ${0.317 \pm 0.138}^{*}$ & \textbf{0.285 $\pm$ 0.138} & $\mathbf{0.486 \pm 0.093}^{*}$ & $\mathbf{0.346 \pm 0.131}^{*}$ & 0.378 $\pm$ 0.081 \\
        \hline
    \end{tabular}    
    \end{adjustbox}
    \label{tab:all_results}
\end{table*}

%% file: tables/all_results_foundation.tex
\begin{table*}[ht]
    \centering
    \caption{
    \rev{Model performance and fairness of the finetuned large pretrained models (foundation models) on the FOG detection task in comparison to the baseline model (A2). 
    \textbf{Bold} text indicates the best value in each column for each dataset. 
     $^{*}$ indicates a statistically significant improvement ($p \leq 0.05$, Wilcoxon Signed Rank test~\cite{woolson2007wilcoxon}) comparing each bias mitigation method to the baseline ($\times$ in Bias mitigation column) within the same dataset.
    Generic activity feature transfer uses a model pretrained on the unlabeled UK Biobank dataset \cite{doherty2017large} and finetuned on the target dataset (in the Dataset column).
    Foundation model feature transfer uses a model pretrained on accelerometer data from NHANES dataset~\cite{centers2019nhanes, ruan2025foundationmodelswearablemovement} and finetuned on the target dataset in the Dataset column.
    }}
    \begin{adjustbox}{max width=\linewidth}
    \small
    \begin{tabular}{cccc|cccc|cccc}
        \hline
         &  & & &  \multicolumn{4}{c|}{Demographic Parity Ratio (DPR)} & \multicolumn{3}{c}{Equalized Odds Ratio (EOR)} &  \makecell{Equality of\\Opportunity\\Difference (EOD)}\\
        Dataset & Model & \makecell{Bias\\mitigation} & F1-score & Sex & Age & \makecell{Disease\\duration} & \makecell{FOG phenotype}  & Sex & Age & \makecell{Disease\\duration} & \makecell{FOG phenotype} \\
        \hline\hline
         
         \multirow{3}{*}[-1em]{DeFOG} & \makecell{A2} & $\times$  & 0.582 $\pm$ 0.010&0.402 $\pm$ 0.068&0.528 $\pm$ 0.076&0.569 $\pm$ 0.077& 0.418 $\pm$ 0.064 & 0.300 $\pm$ 0.054 & 0.320 $\pm$ 0.058 & 0.382 $\pm$ 0.064 & 0.183 $\pm$ 0.027\\
         
         & B1 & \makecell{Generic Activity\\Feature Transfer}  & \textbf{0.659 $\pm$ 0.012} & 0.522 $\pm$ 0.068 & 0.539 $\pm$ 0.063 & 0.592 $\pm$ 0.073 & 0.530 $\pm$ 0.060 & 0.366 $\pm$ 0.058 & 0.395 $\pm$ 0.065 & 0.475 $\pm$ 0.071 & 0.225 $\pm$ 0.033 \\
         
         & B2 & \makecell{Foundation Model\\Feature Transfer} & 0.592 $\pm$ 0.010 &	\textbf{0.639 $\pm$ 0.080}$^*$ &	\textbf{0.657 $\pm$ 0.097}$^*$ &	\textbf{0.621 $\pm$ 0.088}$^*$ & \textbf{0.853 $\pm$ 0.099}$^*$ &	\textbf{0.501 $\pm$ 0.083}$^*$ & \textbf{0.553 $\pm$ 0.093}$^*$ & \textbf{0.509 $\pm$ 0.087}$^*$ & \textbf{0.111 $\pm$ 0.106}\\
        \hline
         \multirow{3}{*}[-1em]{tDCS FOG} & \makecell{A2} & $\times$ & 0.722 $\pm$ 0.019 & 0.614 $\pm$ 0.081&0.598 $\pm$ 0.064&0.591 $\pm$ 0.060 & 0.660 $\pm$ 0.060 & 0.543 $\pm$ 0.083&0.658 $\pm$ 0.054&0.624 $\pm$ 0.053 & 0.228 $\pm$ 0.040\\

        & B1 & \makecell{Generic Activity\\Feature Transfer} & \textbf{0.763 $\pm$ 0.024} & 0.584 $\pm$ 0.086 & 0.567 $\pm$ 0.074 & 0.573 $\pm$ 0.074 & 0.572 $\pm$ 0.069 & 0.523 $\pm$ 0.088 & 0.615 $\pm$ 0.060 & 0.563 $\pm$ 0.057 & 0.321 $\pm$ 0.055 \\
        
        & B2 & \makecell{Foundation Model\\Feature Transfer} & 0.558 $\pm$ 0.020 &	\textbf{0.855 $\pm$ 0.060}$^*$ &	\textbf{0.820 $\pm$ 0.049}$^*$ &	\textbf{0.850 $\pm$ 0.038}$^*$ &	\textbf{0.939 $\pm$ 0.025}$^*$ &	\textbf{0.805 $\pm$ 0.053}$^*$ &	\textbf{0.834 $\pm$ 0.050}$^*$	 & \textbf{0.863 $\pm$ 0.052}$^*$ & \textbf{0.099 $\pm$ 0.027} \\
        \hline
         \multirow{3}{*}[-1em]{De Souza} & \makecell{A2} & $\times$ & \textbf{0.778 $\pm$ 0.033} & 0.464 $\pm$ 0.080&0.423 $\pm$ 0.078&0.481 $\pm$ 0.082 & \textbf{0.786 $\pm$ 0.078} & 0.317 $\pm$ 0.060&0.362 $\pm$ 0.073&0.404 $\pm$ 0.075 & \textbf{0.191 $\pm$ 0.070}\\
        
        & B1 & \makecell{Generic Activity\\Feature Transfer} & 0.721 $\pm$ 0.026&0.422 $\pm$ 0.079&0.511 $\pm$ 0.076&0.491 $\pm$ 0.070 & 0.720 $\pm$ 0.111 & 0.358 $\pm$ 0.068&0.347 $\pm$ 0.075&0.381 $\pm$ 0.068 & 0.213 $\pm$ 0.082 \\
        
        & B2 & \makecell{Foundation Model\\Feature Transfer} &  0.666 $\pm$ 0.032 & \textbf{0.471 $\pm$ 0.079} & \textbf{0.605 $\pm$ 0.079}$^*$ & \textbf{0.624 $\pm$ 0.072}$^*$ & 0.584 $\pm$ 0.113 &	\textbf{0.366 $\pm$ 0.072}	& \textbf{0.405 $\pm$ 0.072} & \textbf{0.467 $\pm$ 0.078} &	0.391 $\pm$ 0.081 \\
        \hline
        \multirow{3}{*}[-1em]{Daphnet} & \makecell{A2} & $\times$ & \textbf{0.644 $\pm$ 0.031} & 0.273 $\pm$ 0.102 & 0.392 $\pm$ 0.095 & 0.345 $\pm$ 0.093 & 0.245 $\pm$ 0.138 & 0.132 $\pm$ 0.072 & 0.323 $\pm$ 0.102 & 0.224 $\pm$ 0.097 & 0.303 $\pm$ 0.094\\
        
        & B1 & \makecell{Generic Activity\\Feature Transfer} & 0.644 $\pm$ 0.032 & \textbf{0.449 $\pm$ 0.160} & 0.411 $\pm$ 0.116 & 0.439 $\pm$ 0.116 & 0.293 $\pm$ 0.125 & 0.132 $\pm$ 0.083 & 0.208 $\pm$ 0.080 & 0.167 $\pm$ 0.068 &  0.267 $\pm$ 0.053 \\
        
        & B2 & \makecell{Foundation Model\\Feature Transfer} & 0.615 $\pm$ 0.022&	${0.412 \pm 0.095}^{*}$ &	\textbf{0.529 $\pm$ 0.098}$^*$	& \textbf{0.512 $\pm$ 0.082}$^*$ &	$\mathbf{0.425 \pm 0.111}^{*}$ & $\mathbf{0.202 \pm 0.078}^{*}$	& $\mathbf{0.371 \pm 0.085}^{*}$ & $\mathbf{0.286 \pm 0.081}^{*}$ & \textbf{0.263 $\pm$ 0.052} \\
        \hline
    \end{tabular} 
    \end{adjustbox}
    \label{tab:all_results_foundation}
\end{table*}

%% file: tables/dataset_protocol_details.tex
\begin{table*}[t]
  \caption{Description of the FOG-provoking tasks, location of study, medication states for the four multi-site datasets used.}
  \label{tab:dataset_protocol_details}
  \begin{adjustbox}{width=\linewidth}
        \small
      \begin{tabular}{c c c p{10cm}}
        \toprule
         Dataset & Location & \makecell{Medication\\States} & \makecell{Study Protocol}\\
        \midrule
        DeFOG & Home & On and off & \makecell[l]{Four meter walk test\\Timed Up \& Go (TUG): series of tasks including rising from a chair, walking a distance of \\ 3 meters, walking back to a chair and sitting down~\cite{podsiadlo1991timed}.\\
        TUG dual task: TUG test with mental subtraction of numbers~\cite{kaggleFOG}.\\
        Hotspot Door task: Walking trial involving opening a door, entering another room,\\and returning to the starting point~\cite{kaggleFOG}. \\Personalized Hostspot Door task: walking through an area in the house identified as\\FOG-provoking~\cite{kaggleFOG}.}\\
        \hline
        tDCS FOG & Clinic & On and off & \makecell[l]{Single task: sit on a chair, rise up \& start walking, turn 360 degrees clockwise and \\counterclockwise, walk 2 meters, open a door, go through, turn 180 degrees,\\and then go back and sit on the chair~\cite{ziegler2010new}.\\
        Dual motor task: perform a single task while carrying a tray with a bottle on \\it~\cite{reches2020using}.\\
        Motor cognitive task: performing the dual motor task while performing serial seven \\subtractions~\cite{reches2020using}.}\\
        \hline
        Daphnet & Clinic & On and off & Daily walking tasks: walking back and forth in a straight line with random 180-degree turns, random walking with a series of initiated stops and 360-degree turns.\\
        \hline
        De Souza & Clinic & Off & Turning tasks: turning in place, alternating 360-degree turns to the right and then to the left, and repeating turning at a self-selected pace for two minutes.\\
      \bottomrule 
    \end{tabular}
    \end{adjustbox}
\end{table*}

%% file: tables/per_dataset_metadata_details.tex
\begin{table*}[t]
\addBox{
\centering
\caption{Clinical metadata availability differs across datasets, which
constrains the attributes available for subgroup fairness evaluation.
$\checkmark$ indicates that the attribute is reported at the subject
level in the dataset's accompanying clinical documentation; ---
indicates that the attribute is not reported, and no inference should
be drawn about whether it was assessed during data collection.
Instruments and scales are given in parentheses.}
\label{tab:metadata}
\small
\begin{tabularx}{\textwidth}{@{}l
  >{\raggedright\arraybackslash}X
  >{\raggedright\arraybackslash}X
  >{\raggedright\arraybackslash}X
  >{\raggedright\arraybackslash}X@{}}
\toprule
Dataset & Medication state & Disease severity & Cognitive status & FOG severity \\
\midrule
Daphnet &
  $\checkmark$ \newline (On or Off medication state provided) &
  $\checkmark$ \newline (Hoehn and Yahr score) &
  --- &
  --- \\
\addlinespace
De Souza &
  $\checkmark$ \newline (Levodopa dosage in mg) &
  $\checkmark$ \newline (Hoehn and Yahr score, total UPDRS-II and UPDRS-III scores) &
  $\checkmark$ \newline (MoCA score) &
  $\checkmark$ \newline (N-FOG-Q) \\
\addlinespace
DeFOG &
  $\checkmark$ \newline (On and Off medication state provided) &
  $\checkmark$ \newline (total UPDRS-III in on and off states) &
  --- &
  $\checkmark$ \newline (N-FOG-Q) \\
\addlinespace
tDCS FOG &
  $\checkmark$ \newline (On and Off medication state provided) &
  $\checkmark$ \newline (total UPDRS-III in on and off states) &
  --- &
  $\checkmark$ \newline (N-FOG-Q) \\
\bottomrule
\end{tabularx}
}
\end{table*}

%% file: supplementary_tables/zero_shot_transfer.tex
\begin{table*}
\centering
\caption{Model performance and fairness on performing zero-shot transfer on A2 (DeepConvLSTM) and A3 (Masked Transformer). 
Mean $\pm$ 95\% confidence interval over 5 runs of leave-one-dataset-out cross-validation (5 random seeds). Each run's value is the mean over its 5 folds. 
In folds where the model detected no FOG episodes, the ratios (DPR, EOR) are undefined and are counted as 0.
}
\label{tab:zero_shot}
\addBox{
\footnotesize
\setlength{\tabcolsep}{3pt}
\renewcommand{\arraystretch}{1.2}
\resizebox{\textwidth}{!}{%
\begin{tabular}{cccc|cccc|cccc}
    \hline
      &    &   &   &  \multicolumn{4}{c|}{Demographic Parity Ratio (DPR)}  &  \multicolumn{3}{c}{Equalized Odds Ratio (EOR)} & \makecell{Equality of \\Opportunity\\Difference (EOD)}\\
    Dataset  &  Model  &  \makecell{Pretrain\\dataset}  &  F1-score  &  Sex  &  Age  &  \makecell{Disease\\duration}  &  \makecell{FOG\\phenotype}  & Sex  &  Age  &  \makecell{Disease\\duration}  &  \makecell{FOG\\phenotype}\\
    \hline\hline
    \multirow{3}{*}{DeFOG}  &  A2  &  \multirow{3}{*}{\makecell{tDCS FOG\\+ Daphnet}}  &  \makecell{0.582\\$\pm$ 0.032} & \makecell{0.445\\$\pm$ 0.152} & \makecell{0.445\\$\pm$ 0.236} & \makecell{0.464\\$\pm$ 0.182} & \makecell{0.088\\$\pm$ 0.049} & \makecell{0.251\\$\pm$ 0.142} & \makecell{0.226\\$\pm$ 0.195} & \makecell{0.285\\$\pm$ 0.104} & \makecell{0.200\\$\pm$ 0.107} \\
      &  A3  &  &  \makecell{0.574\\$\pm$ 0.024} & \makecell{0.330\\$\pm$ 0.051} & \makecell{0.430\\$\pm$ 0.116} & \makecell{0.390\\$\pm$ 0.069} & \makecell{0.073\\$\pm$ 0.025} & \makecell{0.148\\$\pm$ 0.092} & \makecell{0.220\\$\pm$ 0.172} & \makecell{0.203\\$\pm$ 0.047} & \makecell{0.172\\$\pm$ 0.067} \\
    \hline
    \multirow{3}{*}{tDCS FOG}  &  A2  &  \multirow{3}{*}{\makecell{DeFOG\\+ Daphnet}}  &  \makecell{0.411\\$\pm$ 0.005} & \makecell{0.053\\$\pm$ 0.135} & \makecell{0.044\\$\pm$ 0.085} & \makecell{0.042\\$\pm$ 0.116} & \makecell{0.061\\$\pm$ 0.168} & \makecell{0.000\\$\pm$ 0.000} & \makecell{0.000\\$\pm$ 0.000} & \makecell{0.000\\$\pm$ 0.000} & \makecell{0.002\\$\pm$ 0.005} \\
      &  A3  &  &  \makecell{0.450\\$\pm$ 0.059} & \makecell{0.249\\$\pm$ 0.337} & \makecell{0.249\\$\pm$ 0.350} & \makecell{0.218\\$\pm$ 0.313} & \makecell{0.202\\$\pm$ 0.259} & \makecell{0.126\\$\pm$ 0.157} & \makecell{0.182\\$\pm$ 0.288} & \makecell{0.151\\$\pm$ 0.225} & \makecell{0.051\\$\pm$ 0.090} \\
    \hline
    \multirow{3}{*}{Daphnet}  &  A2  &  \multirow{3}{*}{De Souza}  &  \makecell{0.402\\$\pm$ 0.153} & \makecell{0.530\\$\pm$ 0.525} & \makecell{0.460\\$\pm$ 0.520} & \makecell{0.418\\$\pm$ 0.565} & \makecell{0.340\\$\pm$ 0.525} & \makecell{0.389\\$\pm$ 0.480} & \makecell{0.290\\$\pm$ 0.465} & \makecell{0.304\\$\pm$ 0.487} & \makecell{0.321\\$\pm$ 0.530} \\
      &  A3  &  &  \makecell{0.095\\$\pm$ 0.003} & \makecell{0.993\\$\pm$ 0.006} & \makecell{0.996\\$\pm$ 0.005} & \makecell{0.993\\$\pm$ 0.006} & \makecell{0.995\\$\pm$ 0.002} & \makecell{0.837\\$\pm$ 0.113} & \makecell{0.757\\$\pm$ 0.112} & \makecell{0.837\\$\pm$ 0.113} & \makecell{0.995\\$\pm$ 0.003} \\
    \hline
    \multirow{3}{*}{De Souza}  &  A2  &  \multirow{3}{*}{Daphnet}  &  \makecell{0.450\\$\pm$ 0.017} & \makecell{0.000\\$\pm$ 0.000} & \makecell{0.000\\$\pm$ 0.000} & \makecell{0.000\\$\pm$ 0.000} & \makecell{0.002\\$\pm$ 0.006} & \makecell{0.000\\$\pm$ 0.000} & \makecell{0.000\\$\pm$ 0.000} & \makecell{0.000\\$\pm$ 0.000} & \makecell{0.010\\$\pm$ 0.026} \\
      &  A3  &  &  \makecell{0.448\\$\pm$ 0.007} & \makecell{0.022\\$\pm$ 0.061} & \makecell{0.017\\$\pm$ 0.046} & \makecell{0.037\\$\pm$ 0.102} & \makecell{0.029\\$\pm$ 0.074} & \makecell{0.000\\$\pm$ 0.000} & \makecell{0.000\\$\pm$ 0.000} & \makecell{0.000\\$\pm$ 0.000} & \makecell{0.005\\$\pm$ 0.009} \\
    \hline
\end{tabular}}}
\end{table*}

%% file: tables/subject_per_fog_type.tex
\begingroup
\addLong
\setlength{\tabcolsep}{4pt}   
\footnotesize                        
\begin{longtable}{lrrrrrrrr}
\caption{
FOG type distribution per subject. 
Durations are derived from the annotated samples at each dataset's native sampling rate (Daphnet 64~Hz, DeFOG 100~Hz, De Souza and tDCS FOG 128~Hz). 
Tremulous (proportion of total FOG) is undefined (---) for subjects with no annotated freezing.
}\label{tab:subject_fog_type}\\
\toprule
Subject & \makecell{Recordings\\(n)} & \makecell{Akinetic\\(s)} & \makecell{Tremulous\\(s)} & \makecell{Total FOG\\(s)} & \makecell{Total recording\\(s)} & \makecell{Tremulous\\(proportion of\\total FOG)} & \makecell{Proportion of\\Tremulous over\\total recording} & \makecell{Proportion of\\Akinetic over\\total recording}\\
\midrule
\endfirsthead
\multicolumn{6}{l}{\itshape Table \thetable{} (continued)}\\
\toprule
Subject & \makecell{Recordings\\(n)} & \makecell{Akinetic\\(s)} & \makecell{Tremulous\\(s)} & \makecell{Total FOG\\(s)} & \makecell{Total recording\\(s)} & \makecell{Tremulous\\(proportion of\\total FOG)} & \makecell{Proportion of\\Tremulous over\\total recording} & \makecell{Proportion of\\Akinetic over\\total recording}\\
\midrule
\endhead
\midrule
\multicolumn{9}{r}{\itshape continued on next page}\\
\endfoot
\bottomrule
\endlastfoot
\multicolumn{9}{l}{\textbf{Daphnet} (n = 10)}\\
\midrule
S01 & 2 & 1.0 & 103.6 & 104.6 & 1,900.0 & 0.990 & 0.055 & 0.001\\
S02 & 2 & 0.0 & 181.4 & 181.4 & 1,415.0 & 1.000 & 0.128 & 0.000\\
S03 & 3 & 1.5 & 285.0 & 286.6 & 2,010.1 & 0.995 & 0.142 & 0.001\\
S04 & 1 & 0.0 & 0.0 & 0.0 & 2,070.0 & --- & 0.000 & 0.000\\
S05 & 2 & 62.9 & 411.6 & 474.5 & 2,090.1 & 0.867 & 0.197 & 0.030\\
S06 & 2 & 3.1 & 128.4 & 131.5 & 1,990.1 & 0.977 & 0.065 & 0.002\\
S07 & 2 & 4.0 & 77.5 & 81.5 & 1,610.0 & 0.951 & 0.048 & 0.002\\
S08 & 1 & 0.0 & 200.9 & 200.9 & 770.1 & 1.000 & 0.261 & 0.000\\
S09 & 1 & 3.6 & 266.5 & 270.1 & 1,740.1 & 0.987 & 0.153 & 0.002\\
S10 & 1 & 0.0 & 0.0 & 0.0 & 2,230.0 & --- & 0.000 & 0.000\\
\midrule
\textit{Daphnet total} & 17 & 76.1 & 1,654.9 & 1,731.0 & 17,825.5 & 0.956 & 0.093 & 0.004\\
\addlinespace
\multicolumn{9}{l}{\textbf{De Souza} (n = 35)}\\
\midrule
SUB01 & 2 & 0.0 & 231.1 & 231.1 & 240.0 & 1.000 & 0.963 & 0.000\\
SUB02 & 1 & 0.0 & 107.6 & 107.6 & 120.0 & 1.000 & 0.897 & 0.000\\
SUB03 & 3 & 0.0 & 2.7 & 2.7 & 360.0 & 1.000 & 0.008 & 0.000\\
SUB04 & 1 & 0.0 & 17.0 & 17.0 & 120.0 & 1.000 & 0.142 & 0.000\\
SUB05 & 2 & 0.0 & 23.4 & 23.4 & 240.0 & 1.000 & 0.097 & 0.000\\
SUB06 & 2 & 0.0 & 2.3 & 2.3 & 240.0 & 1.000 & 0.009 & 0.000\\
SUB07 & 3 & 0.0 & 355.4 & 355.4 & 360.0 & 1.000 & 0.987 & 0.000\\
SUB08 & 1 & 0.0 & 81.8 & 81.8 & 120.0 & 1.000 & 0.682 & 0.000\\
SUB09 & 2 & 0.0 & 20.9 & 20.9 & 240.0 & 1.000 & 0.087 & 0.000\\
SUB10 & 2 & 0.0 & 0.0 & 0.0 & 240.0 & --- & 0.000 & 0.000\\
SUB11 & 2 & 13.3 & 17.3 & 30.6 & 240.0 & 0.566 & 0.072 & 0.055\\
SUB12 & 1 & 0.0 & 0.0 & 0.0 & 120.0 & --- & 0.000 & 0.000\\
SUB13 & 1 & 0.0 & 0.0 & 0.0 & 120.0 & --- & 0.000 & 0.000\\
SUB14 & 3 & 0.0 & 84.0 & 84.0 & 360.0 & 1.000 & 0.233 & 0.000\\
SUB15 & 1 & 0.0 & 4.0 & 4.0 & 120.0 & 1.000 & 0.033 & 0.000\\
SUB16 & 2 & 0.0 & 39.1 & 39.1 & 240.0 & 1.000 & 0.163 & 0.000\\
SUB17 & 1 & 0.0 & 0.0 & 0.0 & 120.0 & --- & 0.000 & 0.000\\
SUB18 & 3 & 0.0 & 7.0 & 7.0 & 360.0 & 1.000 & 0.020 & 0.000\\
SUB19 & 2 & 0.0 & 236.3 & 236.3 & 240.0 & 1.000 & 0.985 & 0.000\\
SUB20 & 3 & 0.0 & 7.9 & 7.9 & 360.0 & 1.000 & 0.022 & 0.000\\
SUB21 & 3 & 0.0 & 12.4 & 12.4 & 360.0 & 1.000 & 0.034 & 0.000\\
SUB22 & 1 & 0.0 & 0.0 & 0.0 & 120.0 & --- & 0.000 & 0.000\\
SUB23 & 2 & 0.0 & 11.3 & 11.3 & 240.0 & 1.000 & 0.047 & 0.000\\
SUB24 & 1 & 2.8 & 23.2 & 26.1 & 120.0 & 0.891 & 0.194 & 0.024\\
SUB25 & 3 & 0.0 & 0.0 & 0.0 & 360.0 & --- & 0.000 & 0.000\\
SUB26 & 3 & 0.0 & 0.0 & 0.0 & 360.0 & --- & 0.000 & 0.000\\
SUB27 & 3 & 4.2 & 121.4 & 125.6 & 360.0 & 0.966 & 0.337 & 0.012\\
SUB28 & 3 & 0.0 & 0.0 & 0.0 & 360.0 & --- & 0.000 & 0.000\\
SUB29 & 1 & 0.0 & 47.2 & 47.2 & 120.0 & 1.000 & 0.393 & 0.000\\
SUB30 & 2 & 1.5 & 50.0 & 51.4 & 240.0 & 0.972 & 0.208 & 0.006\\
SUB31 & 3 & 0.0 & 17.6 & 17.6 & 360.0 & 1.000 & 0.049 & 0.000\\
SUB32 & 3 & 0.0 & 1.8 & 1.8 & 360.0 & 1.000 & 0.005 & 0.000\\
SUB33 & 1 & 0.0 & 42.4 & 42.4 & 120.0 & 1.000 & 0.353 & 0.000\\
SUB34 & 1 & 25.2 & 0.0 & 25.2 & 120.0 & 0.000 & 0.000 & 0.210\\
SUB35 & 3 & 0.0 & 0.0 & 0.0 & 360.0 & --- & 0.000 & 0.000\\
\midrule
\textit{De Souza total} & 71 & 47.0 & 1,565.0 & 1,612.0 & 8,520.0 & 0.971 & 0.184 & 0.006\\
\addlinespace
\multicolumn{9}{l}{\textbf{DeFOG} (n = 38)}\\
\midrule
00f674 & 4 & 60.8 & 117.6 & 178.3 & 5,316.2 & 0.659 & 0.022 & 0.011\\
040587 & 3 & 30.9 & 287.9 & 318.8 & 3,020.8 & 0.903 & 0.095 & 0.010\\
056372 & 2 & 94.1 & 615.5 & 709.6 & 3,454.7 & 0.867 & 0.178 & 0.027\\
08de77 & 1 & 16.2 & 134.2 & 150.4 & 1,382.4 & 0.892 & 0.097 & 0.012\\
107712 & 2 & 181.1 & 97.8 & 278.9 & 3,461.3 & 0.351 & 0.028 & 0.052\\
12f8d1 & 2 & 7.0 & 1.1 & 8.1 & 2,547.7 & 0.140 & 0.000 & 0.003\\
1a778d & 4 & 241.4 & 429.6 & 671.0 & 7,598.4 & 0.640 & 0.057 & 0.032\\
1d7a0d & 1 & 5.5 & 2.0 & 7.4 & 863.2 & 0.265 & 0.002 & 0.006\\
1fb9cd & 4 & 8.2 & 191.8 & 200.0 & 4,464.7 & 0.959 & 0.043 & 0.002\\
2874c5 & 4 & 114.8 & 100.3 & 215.1 & 5,410.5 & 0.466 & 0.019 & 0.021\\
2a7175 & 2 & 4.2 & 36.5 & 40.7 & 3,112.5 & 0.897 & 0.012 & 0.001\\
387ea0 & 4 & 34.2 & 282.8 & 317.0 & 7,961.2 & 0.892 & 0.036 & 0.004\\
3a90e2 & 4 & 12.2 & 26.7 & 38.9 & 6,634.4 & 0.686 & 0.004 & 0.002\\
3e1d75 & 4 & 30.2 & 22.3 & 52.5 & 6,950.9 & 0.424 & 0.003 & 0.004\\
413532 & 2 & 174.0 & 133.6 & 307.6 & 4,556.4 & 0.434 & 0.029 & 0.038\\
4e6c23 & 2 & 34.7 & 160.4 & 195.0 & 2,620.5 & 0.822 & 0.061 & 0.013\\
575c60 & 2 & 14.9 & 37.4 & 52.3 & 2,639.8 & 0.715 & 0.014 & 0.006\\
58e067 & 2 & 16.9 & 0.7 & 17.7 & 1,337.8 & 0.041 & 0.001 & 0.013\\
5d1cf8 & 2 & 118.2 & 8.8 & 127.0 & 1,851.8 & 0.069 & 0.005 & 0.064\\
5d9cae & 4 & 31.7 & 124.6 & 156.3 & 5,047.5 & 0.797 & 0.025 & 0.006\\
5db717 & 1 & 2.1 & 3.9 & 6.0 & 1,546.4 & 0.652 & 0.003 & 0.001\\
72716b & 2 & 6.9 & 25.1 & 32.0 & 1,117.2 & 0.785 & 0.022 & 0.006\\
7b2e84 & 2 & 3.6 & 3.4 & 7.0 & 2,008.2 & 0.481 & 0.002 & 0.002\\
7da72f & 2 & 0.9 & 5.6 & 6.5 & 2,668.1 & 0.865 & 0.002 & 0.000\\
8c1f5e & 3 & 381.7 & 33.2 & 414.9 & 3,685.3 & 0.080 & 0.009 & 0.104\\
8d43d9 & 2 & 4.0 & 195.8 & 199.8 & 2,434.8 & 0.980 & 0.080 & 0.002\\
a066f4 & 2 & 569.5 & 117.5 & 687.0 & 4,926.0 & 0.171 & 0.024 & 0.116\\
a3a1f9 & 2 & 4.8 & 5.4 & 10.2 & 3,280.6 & 0.532 & 0.002 & 0.001\\
a50c72 & 2 & 55.1 & 222.1 & 277.3 & 3,079.0 & 0.801 & 0.072 & 0.018\\
ab3b2e & 3 & 3.3 & 2.1 & 5.4 & 3,416.0 & 0.395 & 0.001 & 0.001\\
ae2d35 & 3 & 7.4 & 165.1 & 172.5 & 4,042.6 & 0.957 & 0.041 & 0.002\\
bb3387 & 2 & 2.8 & 1.1 & 3.8 & 3,860.3 & 0.273 & 0.000 & 0.001\\
c12ab3 & 1 & 0.0 & 100.1 & 100.1 & 1,079.5 & 1.000 & 0.093 & 0.000\\
c92925 & 2 & 36.9 & 45.0 & 81.9 & 3,226.0 & 0.549 & 0.014 & 0.011\\
d79fa0 & 2 & 62.2 & 59.1 & 121.3 & 3,787.4 & 0.487 & 0.016 & 0.016\\
d89567 & 1 & 36.9 & 5.0 & 42.0 & 1,183.9 & 0.120 & 0.004 & 0.031\\
e86b6e & 2 & 77.5 & 519.0 & 596.5 & 7,556.7 & 0.870 & 0.069 & 0.010\\
f28337 & 2 & 12.1 & 46.1 & 58.2 & 2,126.3 & 0.792 & 0.022 & 0.006\\
\midrule
\textit{DeFOG total} & 91 & 2,498.9 & 4,366.2 & 6,865.1 & 135,257.0 & 0.636 & 0.032 & 0.018\\
\addlinespace
\multicolumn{9}{l}{\textbf{tDCS FOG} (n = 62)}\\
\midrule
02bc69 & 6 & 0.0 & 0.0 & 0.0 & 426.3 & --- & 0.000 & 0.000\\
07285e & 24 & 0.0 & 0.0 & 0.0 & 1,010.6 & --- & 0.000 & 0.000\\
082f01 & 6 & 0.0 & 0.0 & 0.0 & 187.2 & --- & 0.000 & 0.000\\
194d1d & 3 & 16.6 & 1,353.4 & 1,370.0 & 1,458.0 & 0.988 & 0.928 & 0.011\\
19ea47 & 4 & 0.0 & 0.0 & 0.0 & 312.1 & --- & 0.000 & 0.000\\
220a17 & 18 & 16.9 & 57.6 & 74.5 & 985.7 & 0.773 & 0.058 & 0.017\\
231c3b & 24 & 107.2 & 91.1 & 198.2 & 1,126.5 & 0.459 & 0.081 & 0.095\\
242a3e & 24 & 2.2 & 98.2 & 100.4 & 1,298.2 & 0.978 & 0.076 & 0.002\\
24a59d & 12 & 7.4 & 239.2 & 246.6 & 633.3 & 0.970 & 0.378 & 0.012\\
251738 & 24 & 0.0 & 8.2 & 8.2 & 832.5 & 1.000 & 0.010 & 0.000\\
2a39f8 & 14 & 3.2 & 78.4 & 81.5 & 538.7 & 0.961 & 0.145 & 0.006\\
2c98f7 & 18 & 12.2 & 317.8 & 330.0 & 784.3 & 0.963 & 0.405 & 0.015\\
2d57c2 & 20 & 7.5 & 7,720.7 & 7,728.2 & 8,225.1 & 0.999 & 0.939 & 0.001\\
301ada & 12 & 0.0 & 0.0 & 0.0 & 350.7 & --- & 0.000 & 0.000\\
312788 & 9 & 107.3 & 12.0 & 119.3 & 754.4 & 0.101 & 0.016 & 0.142\\
31d269 & 24 & 7.6 & 103.5 & 111.1 & 1,009.4 & 0.932 & 0.103 & 0.007\\
364459 & 18 & 0.0 & 0.0 & 0.0 & 917.8 & --- & 0.000 & 0.000\\
3b2403 & 21 & 0.0 & 0.0 & 0.0 & 1,318.4 & --- & 0.000 & 0.000\\
3b2b7a & 24 & 16.3 & 210.7 & 227.0 & 1,433.4 & 0.928 & 0.147 & 0.011\\
48fd62 & 5 & 111.2 & 232.4 & 343.6 & 616.2 & 0.676 & 0.377 & 0.181\\
4b39ac & 24 & 12.4 & 23.2 & 35.6 & 1,123.7 & 0.653 & 0.021 & 0.011\\
4ba1d3 & 12 & 166.5 & 67.3 & 233.8 & 947.3 & 0.288 & 0.071 & 0.176\\
4bb5d0 & 12 & 66.4 & 98.8 & 165.2 & 876.2 & 0.598 & 0.113 & 0.076\\
4ca9b3 & 18 & 3.5 & 490.3 & 493.8 & 1,081.6 & 0.993 & 0.453 & 0.003\\
4dc2f8 & 18 & 19.6 & 16.0 & 35.6 & 804.0 & 0.449 & 0.020 & 0.024\\
4f13b4 & 12 & 141.1 & 46.5 & 187.6 & 781.4 & 0.248 & 0.060 & 0.181\\
51574c & 2 & 0.0 & 0.0 & 0.0 & 74.2 & --- & 0.000 & 0.000\\
516a67 & 12 & 14.2 & 93.3 & 107.5 & 626.2 & 0.868 & 0.149 & 0.023\\
54ee6e & 11 & 37.9 & 57.0 & 94.9 & 681.0 & 0.601 & 0.084 & 0.056\\
59f492 & 5 & 0.0 & 0.0 & 0.0 & 260.2 & --- & 0.000 & 0.000\\
5c0b8a & 5 & 11.4 & 531.8 & 543.3 & 841.2 & 0.979 & 0.632 & 0.014\\
66341b & 18 & 50.9 & 56.4 & 107.2 & 1,148.2 & 0.526 & 0.049 & 0.044\\
69cc45 & 11 & 0.0 & 0.0 & 0.0 & 548.8 & --- & 0.000 & 0.000\\
6a3e93 & 8 & 0.0 & 0.0 & 0.0 & 227.5 & --- & 0.000 & 0.000\\
743f4e & 5 & 9.7 & 0.0 & 9.7 & 276.0 & 0.000 & 0.000 & 0.035\\
7688c1 & 16 & 0.0 & 0.0 & 0.0 & 605.0 & --- & 0.000 & 0.000\\
79011a & 9 & 10.3 & 1,322.5 & 1,332.8 & 1,570.3 & 0.992 & 0.842 & 0.007\\
7eb666 & 8 & 0.0 & 0.0 & 0.0 & 557.4 & --- & 0.000 & 0.000\\
7fcee9 & 22 & 5.8 & 12.7 & 18.5 & 1,015.6 & 0.689 & 0.013 & 0.006\\
87174c & 9 & 0.0 & 0.0 & 0.0 & 468.1 & --- & 0.000 & 0.000\\
8db7dd & 24 & 6.3 & 188.2 & 194.5 & 1,225.9 & 0.968 & 0.154 & 0.005\\
93f49f & 11 & 54.5 & 20.3 & 74.8 & 744.0 & 0.272 & 0.027 & 0.073\\
9f85da & 3 & 3.3 & 27.5 & 30.9 & 162.0 & 0.892 & 0.170 & 0.020\\
a03db7 & 24 & 0.0 & 329.0 & 329.0 & 1,033.0 & 1.000 & 0.318 & 0.000\\
a80ae4 & 11 & 8.8 & 102.5 & 111.3 & 651.8 & 0.921 & 0.157 & 0.013\\
af82b2 & 12 & 274.0 & 7.0 & 280.9 & 1,053.6 & 0.025 & 0.007 & 0.260\\
b19f77 & 10 & 468.9 & 25.8 & 494.7 & 1,061.7 & 0.052 & 0.024 & 0.442\\
bc3908 & 18 & 19.2 & 44.2 & 63.4 & 1,388.7 & 0.698 & 0.032 & 0.014\\
c7fee4 & 5 & 6.8 & 116.3 & 123.1 & 349.9 & 0.945 & 0.332 & 0.019\\
c85fdf & 4 & 0.0 & 0.0 & 0.0 & 109.0 & --- & 0.000 & 0.000\\
c8e721 & 24 & 7.6 & 41.4 & 49.0 & 954.8 & 0.845 & 0.043 & 0.008\\
c95ab0 & 7 & 0.0 & 0.0 & 0.0 & 359.1 & --- & 0.000 & 0.000\\
d8836b & 16 & 0.0 & 0.0 & 0.0 & 683.5 & --- & 0.000 & 0.000\\
d9312a & 24 & 345.1 & 223.3 & 568.4 & 1,730.5 & 0.393 & 0.129 & 0.199\\
e39bc5 & 4 & 144.6 & 0.0 & 144.6 & 405.0 & 0.000 & 0.000 & 0.357\\
e8919c & 9 & 11.3 & 14.4 & 25.7 & 511.1 & 0.560 & 0.028 & 0.022\\
e9fc55 & 8 & 0.0 & 0.0 & 0.0 & 306.4 & --- & 0.000 & 0.000\\
eeaff0 & 12 & 0.0 & 0.0 & 0.0 & 691.5 & --- & 0.000 & 0.000\\
f2c8aa & 24 & 0.0 & 0.0 & 0.0 & 981.3 & --- & 0.000 & 0.000\\
f62eec & 12 & 33.2 & 205.1 & 238.3 & 992.5 & 0.861 & 0.207 & 0.033\\
f686f0 & 12 & 16.7 & 70.8 & 87.5 & 588.4 & 0.809 & 0.120 & 0.028\\
fa8764 & 12 & 0.0 & 0.0 & 0.0 & 460.7 & --- & 0.000 & 0.000\\
\midrule
\textit{tDCS FOG total} & 833 & 2,365.3 & 14,755.1 & 17,120.4 & 55,177.1 & 0.862 & 0.267 & 0.043\\
\addlinespace
\end{longtable}
\endgroup